\documentclass[12pt]{article}
\usepackage{amsmath}
\usepackage{amssymb}
\usepackage{bm}
\usepackage[dvips]{graphicx}
\usepackage{mathtools}

\newcommand{\llangle}{\langle\!\langle}
\newcommand{\rrangle}{\rangle\!\rangle}

\usepackage{upgreek}
\usepackage{xcolor}
\newcommand{\no}{\nonumber\\}

\begin{document}

\baselineskip=17pt

\begin{titlepage}
\rightline{\tt UT-Komaba/16-5}
\rightline{\tt YITP-16-78}
\rightline\today
\begin{center}

\renewcommand{\thefootnote}{\alph{footnote}}

\vskip 2.5cm
{\Large \bf {Construction of action for heterotic string field theory}}\\
\vskip 0.4cm
{\Large \bf {including the Ramond sector}}
\vskip 1.0cm
{\large {Keiyu Goto${}^{1,}$\footnote{
Present address: \it {Komazawa, Setagaya-ku, Tokyo 154-0012, Japan}} 
and Hiroshi Kunitomo${}^2$}}
\vskip 1.0cm
${}^1${{\it {Institute of Physics, The University of Tokyo}}\\
{\it {Komaba, Meguro-ku, Tokyo 153-8902, Japan}}\\
keiyu.goto.9@gmail.com} 
\vskip 0.5cm
${}^2$
{\it {Center for Gravitational Physics, Yukawa Institute for Theoretical Physics, Kyoto University}}\\
{\it {Kitashirakawa Oiwakecho, Sakyo-ku, Kyoto 606-8502, Japan}}\\
kunitomo@yukawa.kyoto-u.ac.jp

\vskip 2.0cm

\renewcommand{\thefootnote}{\arabic{footnote}}
\setcounter{footnote}{0}

{\bf Abstract}
\end{center}

\noindent
Extending the formulation for open superstring field theory given in arXiv:1508.00366,
we attempt to construct a complete action for heterotic string field theory.
The action is non-polynomial in the Ramond string field $\Psi$, and we construct it
order by order in $\Psi$. Using a dual formulation in which the role of $\eta$ and $Q$
is exchanged, the action is explicitly obtained at the quadratic and quartic  order in $\Psi$  
with the gauge transformations.
\end{titlepage}

\tableofcontents

\newpage

\section{Introduction}
\label{sec:intro}
\setcounter{equation}{0}
There are two complementary formulation of superstring field theories:
the Wess-Zumino-Witten (WZW)-like formulation
\cite{Berkovits:1995ab,Okawa:2004ii,Berkovits:2004xh,Matsunaga:2013mba,Matsunaga:2014wpa},
and the algebraic formulation in terms of the $A_\infty/L_\infty$ structure
\cite{Erler:2013xta,Erler:2014eba}.
The gauge invariant actions for the Neveu-Schwarz (NS) sector
(or the NS-NS sector for the type II superstring)
 in the former can be written in a closed form as WZW-like actions utilizing the large
 Hilbert space. The corresponding actions in the latter are constructed in the small
 Hilbert space using the string 
products satisfying the $A_\infty/L_\infty$ relations, whose explicit
form is obtained by solving a differential equation iteratively.
Now it has been clarified that
two formulations for the open superstring field theory are
interrelated by a partial gauge fixing \cite{Erler:2015rra}.
In spite of this success, it had been difficult to complete the action
so as to include the Ramond (R) string in covariant way for a long time.

However an important progress was recently made in the WZW-like open superstring field theory:
a complete gauge invariant action was constructed \cite{Kunitomo:2015usa}.
Soon afterwards, a similar action realizing a cyclic $A_\infty$ structure
was also constructed \cite{Erler:2016ybs,Konopka:2016grr},
and the relation between two was elucidated \cite{Erler:2016ybs}.
These actions contain both the NS sector and R sector,
describing space-time bosons and fermions, respectively, and completely
specify their interactions.
Therefore we are now in a position to study various off-shell 
aspects of open superstring theory.\footnote{
A closely related approach to the heterotic and type II superstring field theory
has been developed by Sen \cite{Sen:2015hha,Sen:2015uaa}.}
The purpose of this paper is to extend this progress to the case of the heterotic
string field theory.

Although it has been difficult to construct a complete action, including the R sector, 
for heterotic string field theory, the equations of motion was already 
constructed both in the WZW-like formulation \cite{Kunitomo:2013mqa,Kunitomo:2014hba}
and in the algebraic formulation \cite{Erler:2015lya}.
In contrast to those in the open superstring field theory
\cite{Berkovits:2001im,Kunitomo:2015usa},
these equations of motion are nonpolynomial not only in the NS string field
but also in the R string field.
Therefore it is natural to consider that the complete action has also to be 
nonpolynomial in both the NS and R string fields. This is also expected from the simple
consideration on general amplitudes with external fermions.
We need proper interactions, described by the restricted polyhedra
\cite{Saadi:1989tb,Kugo:1989aa}\footnote{See also \cite{Kugo:1989tk,Zwiebach:1992ie}.},
including arbitrary (even) number of R string fields to fill the complete
integration region of the moduli space of such amplitudes.
This makes more difficult to construct a complete gauge invariant action
for the heterotic string field theory.
We attempt to construct a gauge invariant action order by order
in the number of R string, and obtain it up to quartic order.

This paper is organized as follows.
In section 2 we will first briefly summarize the results for the open
superstring field theory given in \cite{Kunitomo:2015usa}.
Several important ingredients to construct the complete action, which
can be straightforwardly extended to the heterotic string field theory,
is introduced.
Then we will explain some basics of the heterotic string field theory
in section 3. We will introduce a dual formulation \cite{Goto:2015pqv} 
exchanging the role of $\eta$ and $Q$, which is useful for our aim.
Section 4 is the main part of the paper.
After introducing R string field in the restricted Hilbert space,
we will attempt to construct a complete action order by order,
first in the coupling constant and then in the R string field.
A gauge invariant action will be obtained at the quadratic and quartic 
order in the R string field, each of which is exact in the NS string
field. In section 5, we will summarize our results, and provide
a few hints to construct a complete action at all order in R string. 
In the Appendix \ref{App A}, we gives an explicit construction of the dual string products. 
The appendix \ref{App B} is added to illustrate how the on-shell physical amplitudes 
are reproduced from the constructed action.

\section{Complete action for open superstring field theory}


In this section we summarize the results given in \cite{Kunitomo:2015usa}
without going into detail. Let us focus on a few key points necessary to construct
a gauge invariant action of the heterotic string field theory.

To begin with, we note that there are two alternative expressions of the WZW-like action
for the NS sector. The original expression given in \cite{Berkovits:2004xh} is
\begin{equation}
 S\ =\ \int^1_0dt\ \langle \tilde{A}_t(t),\eta \tilde{A}_Q(t)\rangle,
\label{NS open action}
\end{equation}
where $\eta$ is the zero mode of $\eta(z)$ and
$\tilde{A}_t$ and $\tilde{A}_Q$ are the \textit{left-invariant forms}
\begin{equation}
 \tilde{A}_t(t)\ =\ g^{-1}(t)\partial_t g(t),\qquad
 \tilde{A}_Q(t)\ =\ g^{-1}(t)Qg(t),
\end{equation}
with $g(t)=e^{\Phi(t)}$.
The NS string field $\Phi$ and its one-parameter extension $\Phi(t)$ are
related through the boundary conditions, $\Phi(1) = \Phi,\ \Phi(0) = 0$.

One can easily see that the action (\ref{NS open action}) can also be written
in the dual form in which the role of $\eta$ and $Q$ is exchanged:
\begin{equation}
S\ =\ - \int^1_0dt\ \langle A_t(t), QA_\eta(t)\rangle,
\label{open dual 1}
\end{equation}
where $A_t(t)$ and $A_\eta(t)$ are the \textit{right-invariant forms}
\begin{equation}
 A_t(t)\ =\ (\partial_t g(t)) g^{-1}(t),\qquad
A_\eta(t)\ =\ (\eta g(t)) g^{-1}(t).
\label{open dual 2}
\end{equation}
As we will see shortly, the latter expression is more suitable for
the complete action, in which the $A_\eta$ plays a special role.
This is not only suitable but essential in the heterotic string field theory 
in which two operators $\eta$ and $Q$ do not appear symmetrically but 
act differently on the closed string products.

In order to include the Ramond sector, an important key point is
to restrict the Ramond string field $\Psi$ by the conditions\footnote{
In this paper we use the same symbol $\Psi$ to denote the string field in the Ramond sector
both for the open superstring and for the heterotic string field. We will not confuse them
since two cases never appear simultaneously.
}
\begin{equation}
\eta\Psi\ =\ 0,\qquad XY\Psi\ =\ \Psi,
\label{const open}
\end{equation}
where $X$ and $Y$ are the picture changing operators acting on states
in the small Hilbert space at picture number $-3/2$ and $-1/2$, respectively:
\begin{equation}
 X\ =\ - \delta(\beta_0)\ G_0 + \delta'(\beta_0)\ b_0,\qquad Y\ =\ -c_0\ \delta'(\gamma_0).
\end{equation}
They satisfy the relations
\begin{equation}
 XYX\ =\ X,\qquad YXY\ =\ Y,
\label{XYX}
\end{equation}
implying the operator $XY$ is a projector:
\begin{equation}
 (XY)^2\ =\ XY.
\end{equation}
The former constraint 
imposes 
 that $\Psi$ is in the small Hilbert space,
and the latter restricts the form of $\Psi$ expanded in the ghost zero-modes as
\begin{equation}
 \Psi\ =\ \phi + (\gamma_0+c_0G)\psi\,.
\end{equation}
This restricted form was already known to be enough to construct consistent 
free superstring field theory\cite{Kazama:1985hd,Kazama:1986cy,Terao:1986ex}.

Note here that the operator $X$ is BRST exact
in the large Hilbert space:
%
\begin{equation}
 X\ =\ \{Q, \Theta(\beta_0)\},
\end{equation}
where $\Theta(x)$ is the Heaviside step function satisfying $\Theta(x)'=\delta(x)$.
More generally, we introduce the following operator $\Xi$ which is more
suitable for use in the large Hilbert space \cite{Erler:2016ybs}:
\begin{equation}
 \Xi\ =\ \xi_0 + (\Theta(\beta_0)\eta\xi_0-\xi_0)P_{-3/2}
+(\xi_0\eta\Theta(\beta_0)-\xi_0)P_{-1/2}\,,
\end{equation}
where $P_n$ is the projector onto states at picture number $n$.
The anti-commutator $\{Q,\Xi\}$ is not identical to $X$,
but equal to $X$ if it acts on a state in the small Hilbert space 
at picture number $-3/2$. In other words, we can use the relation
$X=\{Q,\Xi\}$ on a state in the small Hilbert space at picture number
$-3/2$. Using this $\Xi$, we can define 
an important linear operator $F(t)$ as 
\begin{equation}
 F(t)\ =\ \frac{1}{1+\Xi(D_\eta(t)-\eta)}\
=\ 1 + \sum_{n=1}^\infty (-\,\Xi(D_\eta(t)-\eta))^n,
\label{F open}
\end{equation}
where
\begin{equation}
D_\eta(t) A\ \equiv\ \eta A-A_\eta(t) A +(-1)^AAA_\eta(t),
\end{equation}
on an arbitrary Ramond string field $A$.
This linear operator $F(t)$ satisfies the relation
\begin{equation}
 D_\eta(t)F(t)\ =\ F(t)\eta,
\end{equation}
and thus
the \textit{dressed} Ramond string field $F(t)\Psi$
with the Ramond string field $\Psi$ restricted by the constraints (\ref{const open})
is annihilated by $D_\eta(t)$.

Now a complete gauge invariant action is given by
\begin{equation}
 S\ =\ -\frac{1}{2}\llangle\Psi, YQ\Psi\rrangle
-\int_0^1 dt \langle A_t(t), QA_\eta(t)+(F(t)\Psi)^2\rangle,
\label{comp action open}
\end{equation}
where $\llangle\cdot,\cdot\rrangle$ is the 
BPZ 
inner product
in the small Hilbert space.
We can show that this is invariant
under the gauge transformations \cite{Kunitomo:2015usa}:
\begin{align}
 A_\delta\ =&\ Q\Lambda + D_\eta\Omega + \{F\Psi,F\Xi(\{F\Psi,\Lambda\}-\lambda)\},\\
\delta\Psi\ =&\ Q\lambda + X\eta F\Xi D_\eta(\{F\Psi,\Lambda\}-\lambda),
\end{align}
where $\Lambda$ and $\Omega$ are gauge parameters in the NS sector
and $\lambda$ is a gauge parameter in the Ramond sector satisfying
\begin{equation}
 \eta\lambda\ =\ 0,\qquad XY\lambda\ =\ 0.
\end{equation}

\section{The NS sector of heterotic string field theory}
\label{sec:hetero}
\setcounter{equation}{0}



Next we summarize in this section the known results in the NS sector of the heterotic string 
field theory \cite{Okawa:2004ii,Berkovits:2004xh}. 
In particular, we provide a dual formulation \cite{Goto:2015pqv} which plays a significant role 
when we will include the Ramond sector in the next section.

\subsection{Basic ingredients}


In the heterotic string, the holomorphic sector and anti-holomorphic sector are described 
by superconformal field theory and conformal field theory, respectively. 
The conformal field theory for anti-holomorphic sector consists of the matter
sector with $c=26$, and the reparameterization ghosts,
($\tilde{b}(\bar{z})$,$\tilde{c}(\bar{z})$).
The superconformal field theory for holomorphic sector consists of the matter 
sector with $c=15$, the reparameterization ghosts,
($b(z)$, $c(z)$), 
and the superconformal ghosts, $(\beta(z),\gamma(z))$.
An alternative description using $(\xi(z), \eta(z), \phi(z))$
is known for the superconformal ghost sector  \cite{Friedan:1985ge}, related through the bosonization
relation: 
\begin{align}
\beta(z) =\partial\xi(z)e^{-\phi(z)}, \qquad \gamma(z)= e^{\phi(z)}\eta(z).
\end{align}
Therefore, we can consider two Hilbert spaces for describing the superconformal ghost sector.
One is called the {\it large Hilbert space}, constructed as the Fock space of $\xi(z)$, $\eta(z)$, and $\phi(z)$.
The other called the {\it small Hilbert space} can be
defined as a subspace annihilated by the zero mode of $\eta(z)$,
which is equivalent to the Hilbert space constructed as the Fock space of $\beta(z)$ and $\gamma(z)$.
Note that any $\eta$-exact state belongs to the small Hilbert space due to the nilpotency $\eta^2=0$.



Let $V_1$ and $V_2$ be a pair of heterotic string states 
which satisfy the closed string constraints
\begin{equation}
 b_0V_i\ =\ 0,\qquad L_0^-V_i\ =\ 0,\qquad (i=1,2),
\end{equation}
and belong to the large Hilbert space.
The inner product of them is given by
\begin{align}
\langle  V_1 ,V_2  \rangle = \langle V_1| c_0^- | V_2 \rangle,
\label{BPZ large}
\end{align}
where $\langle V_1|$ denotes the BPZ conjugate of $|V_1\rangle$.
It is non-vanishing when the sums of the ghost number $g$ and the picture number $p$ 
of the two input states are $(g,p)=(4,-1)$.
It satisfies  
\begin{align}
\langle V_1,V_2\rangle =(-1)^{(V_1+1)(V_2+1)}\langle V_2,V_1\rangle,
\end{align}
and
\begin{equation}
 \langle QV_1,V_2\rangle\ =\ (-1)^{V_1}\langle V_1,QV_2\rangle,\qquad
 \langle \eta V_1,V_2\rangle\ =\ (-1)^{V_1}\langle V_1,\eta V_2\rangle.
\label{partial Qeta}
\end{equation}


The interactions of closed strings are described using the string products provided in \cite{Zwiebach:1992ie}
\begin{align}
Q ,\quad [\:\:\cdot\:\:,\:\:\cdot\:\:] ,\quad [\:\:\cdot\:\:,\:\:\cdot\:\:,\:\:\cdot\:\:],\quad\cdots .
\end{align}
The $n$-string product carries ghost number $-2n+3$ (and picture number $0$).
%
%
The string products are graded symmetric upon the interchange of the arguments
\begin{equation}
[V_{\sigma(1)},\dots,V_{\sigma(k)}]\ =\ (-1)^{\sigma(\{V\})} [V_{1},\dots,V_k],
\end{equation}
and cyclic with respect to the inner product: 
\begin{equation}
 \langle  V_1,[V_2,...,V_{n+1}]\rangle\ =\  (-1)^{V_1 +V_2+...+V_n }\langle
 [V_1,V_2,...,V_n],V_{n+1}\rangle\, ,
\end{equation}
Here $\sigma$ denotes the permutation from $\{1,...,n\}$ to $\{{\sigma(1)}, ... ,{\sigma(n)}\}$,
and the factor $(-1)^{\sigma(\{ V \})}$ is the sign factor of 
the permutation from $\{V_1,...,V_n\}$ to $\{V_{\sigma(1)}, ... ,V_{\sigma(n)}\}$.
Defining $[V]=QV$,
the string products satisfy the following relations called {\it the $L_\infty$-relations}:
\begin{equation}
0={\sum_{\sigma}}\sum_{m=1}^n(-1)^{\sigma(\{V\})} \frac{1}{m ! (n-m)!}
[\: [V_{\sigma(1)},\dots,V_{\sigma(m)}],V_{\sigma(m+1)},\dots,V_{\sigma(n)} ]
\label{L inf bos}\, .
\end{equation}
They describe an infinite number of relations, the first few of which is given by 
\begin{align}
0&= Q^2,\\
0&= Q[V_1,V_2] + [QV_1,V_2] +(-1)^{V_1}[V_1, QV_2],\\
0&=Q[V_1,V_2,V_3] + [QV_1,V_2,V_3] +(-1)^{V_1}[V_1, QV_2,V_3]+(-1)^{V_1+V_2}[V_1,V_2,QV_3]\nonumber\\
&\qquad + [[V_1,V_2],V_3] +(-1)^{V_1(V_2+V_3)}[[V_2,V_3], V_1]
+ (-1)^{V_3(V_1+V_2)}[[V_3,V_1],V_2]\ .
\end{align}
The operator 
$\eta$ acts as a derivation on the string products:
\begin{align}
\eta \big[ V_{1} , \dots , V_{n} \big]= \sum_{i=1}^{n-1} (-1)^{1+V_{1} +\dots +V_{k-1}} 
\big[ V_{1} , \dots, \eta V_{k} , \dots , V_{n} \big].
\end{align}

It is useful to introduce new string products $[\cdots]_B$,
\begin{equation}
 [V_1,\cdots,V_n]_B\ \equiv\ \sum_{m=0}^\infty\frac{\kappa^m}{m!}[B^m,V_1,\cdots,V_n],
  \qquad (n\ge1),
\label{shifted products}
\end{equation}
shifted by a Grassmann even NS string field $B$ with ghost number 2 and picture number 0.
If $B$ satisfies the Maurer-Cartan equation
\begin{equation}
 QB + \sum_{n=2}^\infty\frac{\kappa^{n-1}}{n!}[B^n]\ =\ 0,
\label{Maurer Cartan}
\end{equation}
the shifted string products (\ref{shifted products}) satisfy the identical $L_\infty$
relation to (\ref{L inf bos}):
\begin{align}
0={\sum_{\sigma}}\sum_{m=1}^n(-1)^{\sigma(\{V\})} \frac{1}{m ! (n-m)!}
[\: [V_{\sigma(1)},\dots,V_{\sigma(m)}]_B,V_{\sigma(m+1)},\dots,V_{\sigma(n)} ]_B
\label{L inf bos 2}\,.
\end{align}
In particular, setting $n=1$, this relation 
provides
the nilpotency of the shifted BRST charge,
$(Q_B)^2=0$, defined by
\begin{equation}
 Q_BV\ \equiv [V]_B\ =\ QV + \sum_{m=1}^\infty \frac{\kappa^m}{m!}[B^m,V]\,.
\label{shifted Q}
\end{equation}

\subsection{WZW-like action}

On the basis of the WZW-like formulation,
a gauge invariant action for the NS sector
of heterotic string field theory
was provided in \cite{Berkovits:2004xh} by an extension of the Berkovits 
open superstring field theory.
We use a heterotic string field ${{\widetilde{V}}}$ in the large Hilbert space
for the NS sector, which is a Grassmann-odd, and has ghost number $1$ 
and picture number $0$.\footnote{
We put $\ \widetilde{}\ $ on the field $V$ to distinguish it from the field
in the dual formulation introduced in the next subsection.
Two fields $\widetilde{V}$ and $V$ are identical at the leading order in 
the coupling constant $\kappa$, but different at order $\kappa^2$ \cite{Goto:2015pqv}.} 
It also satisfies the closed string constraints
\begin{equation}
b_0^- {{\widetilde{V}}} = 0 \,, \qquad L_0^- {{\widetilde{V}}} = 0 \,.
\end{equation}
We introduce 
a one-parameter extension 
${{\widetilde{V}}} (t)$ 
satisfying ${{\widetilde{V}}} ( 0 ) = 0$ and ${{\widetilde{V}}} (1) = {{\widetilde{V}}} $.
The operators 
$\partial_t$ and $\delta$ as well as $\eta$ 
act as derivations 
on
the string products:
\begin{align}
\mathbb X [ \widetilde{V} _{1}(t) , \dots , \widetilde{V} _{n}(t) ]= 
\sum_{k=1}^{n} (-1)^{\mathbb X( 1+\widetilde{V} _{1}+\dots +\widetilde{V} _{k-1})} 
[ \widetilde{V} _{1}(t) , \dots, \mathbb X \widetilde{V} _{k} (t), \dots , \widetilde{V} _{n}(t) ],
\end{align}
where $\mathbb{X}=\eta$, $\partial_t$ or $\delta$.
%
%
A key ingredient in the WZW-like action is the pure-gauge string field $G(\widetilde{V}(t))$,
which is a Grassmann even functional of $\widetilde{V}(t)$ with ghost number 2 and picture
number 0 satisfying the Maurer-Cartan equation
(\ref{Maurer Cartan}):
\begin{align}
QG(\widetilde{V}) + \sum_{n=2}^\infty \frac{\kappa^{n-1}}{n!}[G(\widetilde{V})^n]\ =\ 0\,.
\label{BOZ PG}
\end{align}
It was shown in \cite{Berkovits:2004xh} that
such a functional $G(\widetilde{V})$ can be obtained
by solving the differential equation
\begin{equation}
 \partial_\tau G(\tau\widetilde{V})\ =\ 
\sum_{m=0}^\infty\frac{\kappa^m}{m!}[G(\tau\widetilde{V})^m,\ \widetilde{V}]\
= 
\ Q_{G(\tau\widetilde{V})}\widetilde{V},
\end{equation}
iteratively with the initial condition, $G=0$ at $\tau=0$, and set $\tau=1$. 

Acting a derivation operator
$\mathbb{X}=\eta$, $\partial_t$, or $\delta$
 on (\ref{BOZ PG}), 
we have
\begin{equation}
 Q_G(\mathbb{X}G)\ =\ 0\,.
\end{equation}
Here $Q_G$ is nilpotent due to (\ref{BOZ PG}).
Since
its cohomology is trivial in the large Hilbert space,
one can find that $\mathbb X G$ is $Q_{G}$-exact
and can define a functional $\Psi _\mathbb X(\widetilde{V})$, 
which we call an associated field, satisfying 
\begin{align}
\mathbb X G(\widetilde{V})\ =\ (-1)^\mathbb X Q_{G(\widetilde{V})} 
\Psi_\mathbb X(\widetilde{V}) .\label{BOZ AF}
\end{align}
We denote $\Psi _t(\widetilde{V})$ for $\Psi _{\partial_t}(\widetilde{V})$ for simplicity.
The associated field $\Psi _\eta(\widetilde{V})$ is Grassmann-even and carries ghost
number $2$ and picture number $-1$. The associated fields $\Psi _t(\widetilde{V})$ and
$\Psi_\delta(\widetilde{V})$ are Grassmann-odd and carry ghost number $1$ and picture number $0$.
These associated fields can also be obtained by iteratively solving the differential equations
\begin{align}
\partial_\tau \Psi _{\mathbb X} (\tau\widetilde{V})\
=\ \mathbb X {{\widetilde{V}}} 
+ \kappa \big[ {{\widetilde{V}}} , \Psi _{\mathbb X} (\tau\widetilde{V}) \big] _{G(\tau\widetilde{V})},
\end{align}
with the initial condition, $\Psi_{\mathbb{X}}=0$ at $\tau=0$, and set $\tau=1$.
Utilizing these functionals $G$ and $\Psi_{\mathbb{X}}$,
a gauge-invariant action can be written in the WZW-like form:
\begin{align}
S_{\scriptscriptstyle{\rm WZW}}
=- \int_0^1 dt \langle \Psi _{t}(t), \eta G(t)\rangle,
\label{action NS}
\end{align}
with $\Psi_{\mathbb{X}}(t)\equiv\Psi_{\mathbb{X}}(\widetilde{V}(t))$ 
and $G(t)\equiv G(\widetilde{V}(t))$.
One can show 
that the variation of the integrand 
becomes a total derivative in $t$
\begin{align} 
\delta \big\langle \Psi _t(t), \eta G(t) \big\rangle
&=\partial_t \big\langle \Psi _\delta(t), \eta G(t) \big\rangle,
\end{align}
and 
thus 
the variation of the action is given by
\begin{align} 
\delta S_{\scriptscriptstyle{\rm WZW}}  
=- \langle \Psi _{\delta}(\widetilde{V})  , \eta G(\widetilde{V})\rangle,
\label{var S NS}
\end{align}
since $\widetilde{V}(0)=0$, and $\Psi_{\mathbb{X}}(0)=G(0)=0$.
From (\ref{var S NS}) we find that
the equation of motion is given by
\begin{align}
\eta G(\widetilde{V})\ =\ 0,
\end{align}
and the action (\ref{action NS}) is invariant under the gauge transformations\footnote{
Note that $\Psi_\delta$ is invertible as a function of $\delta\widetilde{V}$.
See also \cite{Berkovits:2004xh} and \cite{Erler:2015uoa}.
} 
\begin{align}
\Psi _\delta\ =\ Q_{G}\widetilde{\Lambda} + \eta \widetilde{\Omega},
\end{align}
where the gauge parameters $\widetilde{\Lambda}$ and $\widetilde{\Omega}$ 
are Grassmann even with ghost number $0$, 
and carry picture number $0$ and $1$, respectively.
The gauge invariance follows from the nilpotency of $Q_{G}$ and $\eta$,
and one of the relations (\ref{BOZ AF}): $\eta G=-Q_{G} \Psi_\eta$.

\subsection{Dual formulation}


Then we provide a dual formulation for the heterotic string field theory given 
in \cite{Goto:2015pqv}, which is suitable and useful to include the Ramond sector. 
It is dual in the sense that the role of $\eta$ and $Q$ is exchanged,
and natural extension 
of 
 (\ref{open dual 1}) and (\ref{open dual 2})
for the open superstring field theory, 
on the basis of which a complete action in \cite{Kunitomo:2015usa} is constructed.
An explicit construction and more detailed discussion on the dual formulation
is explained in Appendix \ref{App A}.
%


In the dual formulation, an $L_\infty$-structure starting with $\eta$ plays a central role.
Note that, in the case of the open string, a set of products 
$\{\eta, -*\}$ satisfy the $A_\infty$-relations:
$\eta$ is nilpotent, $\eta$ acts as a derivation on 
the star product, and the star product is associative.
As a natural extension of $\{\eta, -*\}$, we introduce a set of products satisfying $L_\infty$-relations,
which we call the {\it dual sting products}:
\begin{align}
\eta ,\quad [\:\:\cdot\:\:,\:\:\cdot\:\:]^\eta ,\quad [\:\:\cdot\:\:,\:\:\cdot\:\:,\:\:\cdot\:\:]^\eta,\quad\cdots .
\label{dual products}
\end{align}
The dual string products are graded commutative upon the interchange of the input string field,
and cyclic:
\begin{align}
[V_{\sigma(1)},\dots,V_{\sigma(k)}]^\eta\ =&\ (-1)^{\sigma(\{V\})} [V_{1},\dots,V_k]^\eta,\\[3pt]
\langle V_1,[V_2,\cdots,V_{n+1}]^\eta\rangle\ =&\ (-1)^{V_1+V_2+\cdots+V_n}\langle [V_1,\cdots,V_n]^\eta,V_{n+1}\rangle\,.
\end{align}
They satisfy the $L_\infty$ relations:
\begin{equation}
 {\sum_{\sigma }}\sum_{k=1}^{n} \frac{1}{k!(n-k)!}(-1)^{\sigma(\{V\})} \big[ [ V_{{\sigma(1)}} , 
\dots , V_{{\sigma (k)}} ]^{\eta } , V_{{\sigma (k+1)}} , \dots , V_{{\sigma (n) }} \big] ^{\eta } = 0 ,
\end{equation}
where we denote $\eta V_i$ as $[ V_i]^\eta$. 
The sign factor
$(-1)^{\sigma(\{ V\})}$ 
is that of the permutation from $\{V_1,...,V_n\}$ to 
$\{V_{\sigma(1)}, ... ,V_{\sigma(n)}\}$.
The $n$-th dual string product carries ghost number $3-2n$ and picture number $n-2$.
We also require that
the BRST operator $Q$ acts as a derivation 
on 
the dual string products:
\begin{align}
Q \big[ V_{1} , \dots , V_{n} \big] ^{\eta } + \sum_{k=1}^{n} (-1)^{V_{1} +\dots 
+V_{k-1}} \big[ V_{1} , \dots, Q V_{k} , \dots , V_{n} \big] ^{\eta } = 0 .
\end{align}
We can actually construct such dual string products from the well-known string products,
$\xi_0$ and the picture changing operator $X_0=\{Q,\xi_0\}$, details of which is given 
in \cite{Goto:2015pqv} or Appendix~\ref{App A}. 
For later use, we introduce a one parameter extension 
$V (t)$ satisfying $V (0) = 0$ and $V (1) = V$.
The operator $\mathbb{X}=Q$, $\partial_t$, or $\delta$
acts as a derivation on the dual string products:
\begin{align}
\mathbb X [ V _{1} , \dots , V _{n} ]^\eta= 
\sum_{k=1}^{n} (-1)^{\mathbb X( 1+V _{1} +\dots +V _{k-1})} [ V _{1} , \dots, \mathbb X V _{k} , \dots , V _{n} ]^\eta\,.
\end{align}



Utilizing the dual string products, we can provide an alternative gauge-invariant action 
in the dual manner to that for the WZW-like action reviewed in the previous subsection.
In the dual formulation, we denote the NS string field as $V$,
which is a Grassmann-odd state in the large Hilbert space
with ghost number $1$ and picture number $0$. It satisfies the closed string constraint:
\begin{equation}
b_0^- V = 0 \,, \qquad L_0^- V = 0 \,.
\end{equation}
%
Pure-gauge string field ${G_\eta}(V)$ in the dual formulation is defined as
a functional of $V(t)$ with ghost number $2$ and picture number $-1$
satisfying the Maurer-Cartan equation dual to (\ref{BOZ PG}):
\begin{align}
0\ =\ \eta {G_\eta}(V) 
+ \sum_{n=2}^\infty\frac{\kappa^{n-1}}{n!} [G_\eta(V)^n]^\eta.
\label{MC}
\end{align}
As with $G$ satisfying (\ref{BOZ PG}), $G_\eta$ can be obtained by solving the
differential equation
\begin{align}
\partial_\tau {G_\eta}(\tau V)\ =\ \sum_{m=0}^\infty\frac{\kappa^m}{m!}[G_\eta(\tau V)^m, V]^\eta,
\end{align}
iteratively with $G_\eta(0)=0$, and 
setting 
$\tau=1$.
%
We define the shifted products of dual string products as
\begin{align}
[V_1,V_2, \cdots,V_n \:]^\eta_{{G_\eta}}
\equiv\sum_{m=0}^\infty \frac{\kappa^m}{m!}
[(G_\eta)^m,V_1,V_2,\cdots,V_n]^\eta,
\qquad (n\ge1),
\end{align}
which are also graded commutative and cyclic.
In particular, it is useful to define the shifted $\eta$-operator  $D_\eta$
as the shifted one-string product 
$[\ \cdot\ ]^\eta_{G_\eta}$:
\begin{align}
\label{shifted L-infty}
D_{\eta } V\
\equiv\ [V]^\eta_{{G_\eta}}\
=&\ \sum_{m=0}^\infty \frac{\kappa^m}{m!}
[(G_\eta)^m,V]^\eta
\nonumber\\
=&\ \eta V + \sum_{m=1}^\infty \frac{\kappa^m}{m!}
[(G_\eta)^m,V]^\eta
\,.
\end{align}
The shifted dual string products satisfy the $L_\infty$ relation:
\begin{equation}
 {\sum_{\sigma }}\sum_{k=1}^{n} \frac{1}{k!(n-k)!}(-1)^{\sigma(\{V\})} \big[ [ V_{{\sigma(1)}} , 
\dots , V_{{\sigma (k)}} ]^{\eta }_{G_\eta} , 
V_{{\sigma (k+1)}} , \dots , V_{{\sigma (n) }} \big] ^{\eta }_{G_\eta}\ =\ 0\,.
\end{equation}
Their lowest two relations represent that $D_\eta$ is nilpotent and acts as a derivation
on the dual shifted two string products:
\begin{align}
(D_\eta)^2 V_1\ =&\ 0,\\
D_\eta [V_1,V_2]^\eta_{{G_\eta}}\ =&\
-[D_\eta V_1,V_2]^\eta_{{G_\eta}}-(-1)^{V_1} [V_1,D_\eta V_2]^\eta_{{G_\eta}}\,.
\label{Deta derivation}
\end{align}
The operator 
$\mathbb{X}=Q$, $\partial_t$, or $\delta$
acts on the shifted dual products as
\begin{align}
\mathbb X [ V _{1} , \dots , V _{n} ]^\eta_{{G_\eta}}\ =&\ \sum_{k=1}^{n} 
(-1)^{\mathbb X( V _{1} +\dots +V _{k-1}+1)} [ V _{1} , \dots, \mathbb X V _{k} , \dots , V _{n} ]^\eta_{{G_\eta}}
\nonumber\\
&
+(-1)^\mathbb X \kappa[\mathbb X {G_\eta} ,V _{1} , \dots , V _{n} ]^\eta_{{G_\eta}}.
\end{align}
In particular,
\begin{align}
\mathbb X D_\eta V_1  &=
(-1)^{\mathbb X} D_\eta\mathbb X  V _1 +
 (-1)^\mathbb X\kappa[\mathbb X {G_\eta}, V _1]^\eta_{{G_\eta}}.
\label{Deta X}
\end{align}
%
%
Acting $\mathbb{X}$ on the Maurer-Cartan equation (\ref{MC}) we have
\begin{equation}
 D_\eta\mathbb{X}G_\eta(V)\ =\ 0\,.
\end{equation}
Thus, since $D_\eta$ 
cohomology is trivial,  
$\mathbb X {{G_\eta}}(V)$ is $D_\eta$-exact 
and can be written as
\begin{align}
\mathbb X {{G_\eta}}(V)\  =\ (-1)^\mathbb X  D_\eta { B} _\mathbb X(V),
\label{def BX}
\end{align}
by introducing associated fields $B_{\mathbb{X}}(V)$.
The associated field ${ B} _Q$ is Grassmann-even and carries ghost number $2$  and picture number $0$,
and ${ B} _t\ (\equiv B_{\partial_t})$ and ${ B} _\delta$ are Grassmann-odd 
and carry ghost number $1$ and picture number $0$.
They are obtained by solving the differential equations
\begin{align}
\partial_\tau { B} _{\mathbb X} (\tau V)
= \mathbb X {{{V}}} +  
\kappa   
\big[ {{{V}}}, { B} _{\mathbb X} (\tau V) \big]^\eta
 _{{{G_\eta}}(\tau V)}\,,
\end{align}
iteratively with $B_{\mathbb{X}}=0$ at $\tau=0$, and then set $\tau=1$.
By multiplying
$0=\mathbb{X}\mathbb{Y}-(-1)^{\mathbb{X}\mathbb{Y}}\mathbb{Y}\mathbb{X}$ to $G_\eta$
for $\mathbb{X}, \mathbb{Y}=Q$, $\partial_t$, or $\delta$, 
we can show that the identity
\begin{align}
D_\eta\big( \mathbb X { B} _\mathbb Y
-(-1)^{\mathbb X \mathbb Y}\mathbb Y { B} _\mathbb X
-(-1)^{\mathbb X}\kappa[{ B} _\mathbb X,{ B} _\mathbb Y]^\eta_{{{G_\eta}}}\big)\ =\ 0\,,
\label{dual FXY}
\end{align}
%
which is useful later, holds using (\ref{def BX}), (\ref{Deta X}), and (\ref{Deta derivation}).
We can also show
\begin{equation}
 \langle D_\eta V_1,V_2\rangle\ =\ (-1)^{V_1}\langle V_1,D_\eta V_2\rangle,
\label{partial Deta}
\end{equation}
from the definition (\ref{shifted L-infty}).

An alternative gauge invariant action, which we call the dual WZW-like action, is given
using these functionals $G_\eta(V)$ and $B_t(V)$ by
\begin{align}
S\ =\ \int^1_0 dt\ \langle B_{t}(t), Q {G_\eta}(t) \rangle\,,
\label{dual action NS}
\end{align}
with $B_t(t)\equiv B_t(V(t))$ and $G_\eta(t)\equiv G_\eta(V(t))$.
The variation of the action can be calculated as 
\begin{align}
\delta S
=\langle B_\delta(V), Q {G_\eta(V)} \rangle\,,
\label{eom 0}
\end{align}
in a completely parallel manner with the original formulation in \cite{Berkovits:2004xh}.
Thus the equation of motion is given by
\begin{align}
Q{G_\eta}(V)\ =\ 0\,,
\end{align}
and the action is invariant under the gauge transformation 
\begin{align}
B_\delta= Q\Lambda + D_\eta \Omega  \,.
\label{gauge tf NS}
\end{align}
The gauge parameters ${\Lambda}$ and ${\Omega}$ having ghost number $0$
carry picture number $0$ and $1$, respectively.
The gauge invariance follows from the nilpotency $(D_\eta)^2=Q^2=0$, and $Q {G_\eta} =- D_\eta B_Q $.


Another important property of the dual string products is their 
\textit{Q-exactness}.
%
%
%
%
The dual $n$-string products for $n\ge3$
themselves written as a BRST variation of some products $(\cdots)^\eta$
which we call the {\it dual gauge products}:
\begin{align}
[V_1, \cdots ,  V_n]^\eta  = Q  (V_1, \cdots , V_n)^\eta 
- \sum_{k=1}^n (-1)^{V_1 + \cdots + V_{k-1}} (V_1, \cdots , Q V_k,\cdots , V_n)^\eta,\quad
(n\ge3)\,,
\label{Q gauge product}
\end{align}
which is consistent with the fact that $Q$ acts as a derivation on the dual string products.
Since the dual string products are Grassmann odd, the dual gauge products are Grassmann-even.
The $n$-th dual gauge product carries ghost number $-2n+2$ and picture number $n-2$, and is commutative and cyclic:
\begin{align}
(V_{\sigma(1)},\cdots,V_{\sigma(n)})^\eta &
=(-1)^{\sigma(\{ V\})} ( V_{1},\cdots, V_n)^\eta,\\
\langle  V_1, ( V_2,\cdots, V_{n+1})^\eta\rangle
&=(-1)^{ V_2+\cdots + V_n +1}\langle ( V_1,\cdots, V_n)^\eta , V_{n+1}\rangle\,,
\end{align}
where $(-1)^{\sigma(\{ V\})}$ is the sign factor of the permutation from $\{V_1,..., V_n\}$ 
to $\{ V_{\sigma(1)}, ... , V_{\sigma(n)}\}$.
The operator $\mathbb{X}=\partial_t$, or $\delta$ acts as a derivation
also on this product,
\begin{align}
\mathbb X (V_1, \cdots, V_n)^\eta\ =\
\sum_{i=1}^{n-1}(-1)^{(V_1+\cdots+V_{i-1})}(V_1,\cdots,\mathbb{X}V_i,\cdots,V_n)^\eta\,.
\end{align}
It is useful again to 
define the {\it shifted dual gauge products} $( \cdots )^\eta_{G_\eta}$
by
\begin{equation}
( V_1,\cdots, V_n)^\eta_{G_\eta}\
=\ \sum_{m=0}^\infty\frac{\kappa^m}{m!}((G_\eta)^m,V_1,\cdots,V_n)^\eta\,,\qquad (n\ge3).
\end{equation}
Note that the $n$-th shifted dual product contains all the dual products higher than $n$.
The shifted dual products are cyclic, which follows from the cyclicity of the dual products:
\begin{align}
\langle V_1, ( V_2,\cdots, V_{n+1})^\eta_{G_\eta}\rangle
=(-1)^{ V_2+\cdots+ V_n +1}\langle ( V_1,\cdots, V_n)^\eta_{G_\eta} , V_{n+1}\rangle.
\end{align}
They are related to the shifted dual products $[V_1, ... , V_n]^\eta_{{G_\eta}}$ as follows.
\begin{align}
[ V_1, \cdots ,  V_n]^\eta_{{G_\eta}}
=&\ \sum_{m=0}^\infty \frac{\kappa^m}{m!} 
[(G_\eta)^m , \cdots,  V_1 , \cdots ,  V_n]^\eta\no
=&\ \sum_{m=0}^\infty \frac{\kappa^m}{m!} 
\Big(Q ((G_\eta)^m, V_1, \cdots ,  V_n)^\eta
-m ((G_\eta)^{m-1},Q{G_\eta},  V_1, \cdots , V_n)^\eta \no
&\hspace{60pt}
-\sum_{k=1}^n (-1)^{V_1+...+V_{k-1}}((G_\eta)^m, V_1, \cdots , Q V_k, \cdots , V_n)^\eta \Big)\no
=&\
Q(V_1, \cdots , V_n)^\eta_{G_\eta}
-\sum_{k=1}^n (-1)^{ V_1+\cdots+ V_{k-1}}( V_1, \cdots ,
Q V_k, \cdots ,  V_n)^\eta_{G_\eta}\nonumber\\
&
-\kappa(Q{G_\eta},  V_1, \cdots , V_n)^\eta_{G_\eta}\,.\label{SDGP}
\end{align}
Due to the shift, the operator $\mathbb{X}=\partial_t$, or $\delta$ does not act as a derivation 
on the shifted dual gauge product but satisfies the relation
\begin{align}
\mathbb X ( V_1, \cdots ,  V_n)^\eta_{G_\eta} 
=\sum_{k=1}^n  ( V_1, \cdots , \mathbb X  V_k,\cdots,  V_n)^\eta_{G_\eta} 
+\kappa (\mathbb X{G_\eta},  V_1, \cdots ,  V_n)^\eta_{G_\eta}\,.
\end{align}



\section{Inclusion of the Ramond sector}
\setcounter{equation}{0}

Now let us include the Ramond sector.
In this section, after introducing the Ramond string field constrained into the restricted
Hilbert space, we attempt to construct a gauge invariant action
order by order in the coupling constant $\kappa$. 
The result can easily be extended to the full order for the part of
the action quadratic in fermion, which has the form of a natural extension 
of the complete action for the open superstring field theory.
In the heterotic string case, however, it is not gauge invariant, and
necessary to include the interactions containing arbitrary even number of
Ramond string fields. We determine the quartic term explicitly.

\subsection{Ramond string field and restricted Hilbert space}

Following the case of the open superstring field theory \cite{Kunitomo:2015usa}, 
we introduce a string field $\Psi$ constrained in the restricted Hilbert space,
\begin{equation}
 \eta\Psi\ =\ 0,\qquad XY\Psi\ =\ \Psi,
\label{constR1}
\end{equation}
for the Ramond sector.
It is a Grassmann even state with the ghost number $2$ and the picture number $-1/2$, 
and satisfies the closed string constraint
\begin{equation}
 b_0^-\Psi\ =\ L_0^-\Psi\ =\ 0.
\label{constR2}
\end{equation}
%
%
The picture changing operators $X$ and $Y$ are defined by
\begin{equation}
 X\ =\ -\delta(\beta_0)G_0 + \delta'(\beta_0) b_0,\qquad
 Y\ =\ -2c_0^+\delta'(\gamma_0),
\label{PCO}
\end{equation}
which act on states in the small Hilbert space with the picture 
number $-3/2$ and $-1/2$, respectively. These operators 
are inverse each other in the sense that they satisfy
\begin{equation}
 XYX\ =\  X,\qquad YXY\ =\ Y\,, 
\end{equation}
which make the operator $XY$ a projector:
\begin{equation}
 (XY)^2\ =\ XY.
\end{equation}
In addition, $X$ is commutative with the BRST charge $Q$, $[Q, X]\ =\ 0$.
These are enough to guarantee the compatibility of the restriction with
the BRST cohomology, that is,
if $XY\Psi_1=\Psi_1$ then $XYQ\Psi_1=Q\Psi_1$, which can be shown as
\begin{equation}
 XYQ\Psi_1\ =\ XYQXY\Psi_1\ =\ XYXQY\Psi_1\ =\ XQY\Psi_1\ =\ QXY\Psi_1\ =\ Q\Psi_1.
\end{equation}
The operator $Y$ is chosen to be commutative with $b_0^-$ so that all 
the constraints in (\ref{constR1})  and (\ref{constR2}) are consistent.
Expanding the ghost zero modes,  
the restricted Ramond string field has the form
\begin{equation}
 \Psi\ =\ \phi + (\gamma_0+2c_0^+G)\psi,
\end{equation}
with
\begin{equation}
 L_0^-\phi\ =\ b_0^\pm\phi\ =\ \beta_0\phi\ =\ 0,\qquad
L_0^-\psi\ =\ b_0^\pm\psi\ =\ \beta_0\psi\ =\ 0,
\end{equation}
where $G=G_0+2b_0\gamma_0$.

The appropriate inner product in the restricted Hilbert
space is given by\footnote{
We assume that both $\Psi_1$ and $\Psi_2$ have picture number
$-1/2$, which is enough to define the action (\ref{Rkin}).} 
\begin{equation}
 \llangle\Psi_1,Y\Psi_2\rrangle,
\end{equation}
using the BPZ inner product $\llangle A, B \rrangle$ 
in the small Hilbert space restricted by (\ref{constR2}):
\begin{equation}
 \llangle A, B\rrangle\ =\ \llangle A|c_0^-|B\rrangle.
\end{equation}
The state $\llangle A|$ is the BPZ conjugate of $|A\rrangle$.
Using this inner product we take the free action for the Ramond 
sector to be
\begin{equation}
S_0\ =\ -\, \frac{1}{2}\llangle\Psi, YQ\Psi\rrangle,
\label{Rkin}
\end{equation}
which is invariant under the gauge transformation
\begin{equation}
\delta\Psi\ =\ Q\lambda\,.
\end{equation}
The gauge parameter $\lambda$ also
satisfies the same constraints as $\Psi$:
\begin{equation}
 b_0^-\lambda\ =\ L_0^-\lambda\ =\ \eta\lambda\ =\ 0,\qquad
XY\lambda\ =\ \lambda.
\end{equation}
The properties of string fields and gauge parameters are summarized
in Table \ref{table:string-fields}.

\begin{table}[t]
\begin{center}
{\renewcommand\arraystretch{1.5}
\begin{tabular}{|c||c|c||c|c|c|}
\hline
field & $V$ & $\Psi$ & $\Lambda$ & $\Omega$ & $\lambda$ \\
\hline
Grassmann & odd & even & even & even & odd \\
\hline
$( \boldsymbol{g}, \boldsymbol{p} )$ & $(1,0)$ & $(2,-1/2)$ & $(0,0)$ & $(0,1)$ & $(1,-1/2)$ \\
\hline
\end{tabular}
}
\caption{Properties of the string fields and the gauge parameters with
the ghost number $\boldsymbol{g}$ and the picture number $\boldsymbol{p}$.
}
\label{table:string-fields}
\end{center}
\end{table}

In order to prove the gauge invariance of the action, we need to note that
the operator $X$ is BRST trivial in the large Hilbert space \cite{Preitschopf:1989fc}:
\begin{equation}
 X\ =\ \{Q,\Theta(\beta_0)\},
\end{equation}
with the Heaviside step function $\Theta(x)$.
More general operator $\Xi$ suitable in the large Hilbert space is defined by
\cite{Erler:2016ybs}
\begin{equation}
 \Xi\ =\ \xi + (\Theta(\beta_0)\eta\xi-\xi)P_{-3/2}
+(\xi\eta\Theta(\beta_0)-\xi)P_{-1/2}\,,
\end{equation}
where $P_n$ is a projector onto the states 
with picture number $n$. We can show that this operator $\Xi$ is BPZ even
for the BPZ inner product in the large Hilbert space (\ref{BPZ large}):
\begin{equation}
\langle\Xi V_1, V_2\rangle\ =\ (-1)^{V_1+1}\langle V_1, \Xi V_2\rangle.
\label{BPZ Xi}
\end{equation}
Then we generalize the operator $X$ to the one given by 
\begin{equation}
 X\ =\ \{Q, \Xi\}\,,
\end{equation}
which is identical to $X$ in (\ref{PCO})
on states in the small Hilbert space with the picture number $-3/2$.
Hereafter we only use the new operator, so we denote it by the same symbol $X$
for simplicity.
The operator $X$ is BPZ even with respect to the inner product
in the small Hilbert space
: 
\begin{equation}
 \llangle XV_1, V_2\rrangle\ =\ \llangle V_1, XV_2\rrangle.
\end{equation}

\subsection{Perturbative construction}


A complete action including interactions between the NS sector and 
the Ramond sector can be expanded in powers of fermion:
\begin{equation}
 S\ =\ \sum_{n=0}^\infty S^{(2n)}.
\end{equation}
For the NS sector, $S_{NS}\equiv S^{(0)}$,
we adopt the dual WZW-like action
defined in (\ref{dual action NS}).
The remaining part, 
$S_R\equiv\sum_{n=1}^\infty S^{(2n)}$, contains the kinetic term of
the Ramond sector (\ref{Rkin}) 
and interaction terms between two sectors.
We can further expand the action in the coupling constant $\kappa$:
\begin{align}
 S_{NS}\ =&\ S^{(0)}_0 + \kappa S^{(0)}_1 + \kappa^2 S^{(0)}_2 + O(\kappa^3),\\
 S_{R}\ =&\ S^{(2)}_0 + \kappa S^{(2)}_1 + \kappa^2 (S^{(2)}_2 + S^{(4)}_2) 
+ O(\kappa^3).
\end{align}
The gauge transformations can also be expanded in $\kappa$ as
\begin{align}
 \delta_\Lambda V\ =&\ \delta_\Lambda V_{0}^{(0)} + \kappa \delta_\Lambda V_1^{(0)}
+ \kappa^2 \left(\delta_\Lambda V_2^{(0)} + \delta_\Lambda V^{(2)}_2\right) + O(\kappa^3),\\
 \delta_\Lambda \Psi\ =&\ \delta_\Lambda \Psi^{(1)}_0 + \kappa \delta_\Lambda \Psi^{(1)}_1
+ \kappa^2 \delta_\Lambda \Psi^{(1)}_2 + O(\kappa^3),
\end{align}
with
\begin{equation}
 \delta_\Lambda V^{(0)}_0\ =\ Q\Lambda,\qquad 
\delta_\Lambda\Psi^{(1)}_0\ =\ 0,
\label{tf Lambda 0}
\end{equation}
where $\Lambda$ is a gauge parameter in the NS sector,
\begin{align}
 \delta_\Omega V\ =&\ \delta_\Omega V^{(0)}_0 + \kappa \delta_\Omega V^{(0)}_1
+ \kappa^2 \left(\delta_\Omega V^{(0)}_2 + \delta_\Omega V^{(2)}_2\right) + O(\kappa^3),\\
 \delta_\Omega \Psi\ =&\ \delta_\Omega \Psi^{(1)}_0 + \kappa \delta_\Omega \Psi^{(1)}_1
+ \kappa^2 \delta_\Omega \Psi^{(1)}_2 + O(\kappa^3),
\end{align}
with
\begin{equation}
 \delta_\Omega V^{(0)}_0\ =\ \eta\Omega,\qquad \delta_\Omega\Psi^{(1)}_0\ =\ 0,
\label{tf Omega 0}
\end{equation}
where $\Omega$ is another gauge parameter in the NS sector, and
\begin{align}
 \delta_\lambda V\ =&\ \delta_\lambda V^{(2)}_0 + 
\kappa \delta_\lambda V^{(2)}_1
+ \kappa^2 \delta_\lambda V^{(2)}_2 + O(\kappa^3),\\
 \delta_\lambda \Psi\ =&\ \delta_\lambda \Psi^{(1)}_0 + \kappa \delta_\lambda \Psi^{(1)}_1
+ \kappa^2 \left(\delta_\lambda \Psi^{(1)}_2 + \delta_\lambda\Psi^{(3)}_2\right) + O(\kappa^3),
\end{align}
with
\begin{equation}
\delta_\lambda V^{(2)}_0\ =\ 0,\qquad \delta_\lambda\Psi^{(1)}_0\ =\ Q\lambda,
\label{tf lambda 0}
\end{equation}
where $\lambda$ is a gauge parameter in the Ramond sector.
The number in the parentheses in the superscript of gauge transformations 
denotes the number of fields in the Ramond sector included. 
Starting from the kinetic terms
\begin{align}
 S^{(0)}_0\ =&\ \frac{1}{2}\langle V,Q\eta V\rangle,\\
 S^{(2)}_0\ =&\ -\frac{1}{2}\llangle \Psi, YQ\Psi\rrangle,
\end{align}
let us first attempt to 
construct the action $S$, and simultaneously the gauge transformations, 
order by order in $\kappa$ by requiring the gauge invariance.

For the NS sector, we can obtain the cubic and quartic terms simply by expanding 
the action (\ref{dual action NS}):
\begin{subequations} \label{dual NS action obo}
\begin{align}
 S^{(0)}_1\ =&\ \frac{1}{3!}\langle V, Q[V, \eta V]^\eta\rangle,\\ 
 S^{(0)}_2\ =&\ \frac{1}{4!}\langle V, Q[V, (\eta V)^2]^\eta\rangle
 + \frac{1}{4!}\langle V, Q[V, [V, \eta V]^\eta]^\eta\rangle.
\end{align}
 \end{subequations}
Expanding the gauge transformation (\ref{gauge tf NS}), one can also obtain 
\begin{alignat}{3}
 \delta_\Lambda V^{(0)}_1\ =&\ -\frac{1}{2}[V,Q\Lambda]^\eta, &\qquad
 \delta_\Lambda V^{(0)}_2\ =&\ -\frac{1}{3}[V,\eta V, Q\Lambda]^\eta
+\frac{1}{12}[V,[V,Q\Lambda]],\\
 \delta_\Omega V^{(0)}_1\ =&\ \frac{1}{2}[V,\eta\Omega]^\eta,&\qquad
\delta_\Omega V^{(0)}_2\ =&\ \frac{1}{3!}[V,\eta V,\eta\Omega]^\eta
+\frac{1}{12}[V,[V,\eta\Omega]],
\end{alignat}
which keep the action (\ref{dual NS action obo}) invariant at each order in $\kappa$:
\begin{alignat}{3}
(\delta_\Lambda)^{(0)}_0 S^{(0)}_1 + (\delta_\Lambda)^{(0)}_1 S^{(0)}_0\ =&\ 0,&\qquad
(\delta_\Lambda)^{(0)}_0 S^{(0)}_2 + (\delta_\Lambda)^{(0)}_1S^{(0)}_1 
+ (\delta_\Lambda)^{(0)}_2 S^{(0)}_0\ =&\ 0,\\     
(\delta_\Omega)^{(0)}_0 S^{(0)}_1 + (\delta_\Omega)^{(0)}_1 S^{(0)}_0\ =&\ 0,&\qquad
(\delta_\Omega)^{(0)}_0 S^{(0)}_2 + (\delta_\Omega)^{(0)}_1 S^{(0)}_1 
+ (\delta_\Omega)^{(0)}_2 S^{(0)}_0\ =&\ 0.
\end{alignat}
The number in the parentheses in the superscript of $\delta$ denotes
the difference of the number 
of 
the Ramond field after and before the transformation:
($\#$ of R fields after transformation) $-$ ($\#$ of R fields before transformation).

\subsubsection{Cubic interaction in $S_R$}

Let us consider the cubic interaction in the Ramond action $S_R$.
We start from a natural candidate of cubic interaction term given by
\begin{equation}
 S^{(2)}_1\ =\ \alpha_1\langle\Psi, [V,\Psi]^\eta\rangle
 ,
\end{equation}
with a constant $\alpha_1$ to be determined, 
and find $\delta V_1^{(2)} ( =\delta_1^{(2)} V)$ and $\delta \Psi_1^{(1)} (=
\delta^{(0)}_1\Psi)$ requiring the gauge invariances in this order
\begin{equation}
\delta^{(0)}_0 S^{(2)}_1
+\delta^{(0)}_1 S^{(2)}_0
+\delta^{(2)}_1 S^{(0)}_0 =0\,.
\end{equation}
Note that $\delta^{(2)}_0$ does not appear 
and that $\delta^{(2)}_1$ appear only for the transformation with $\lambda$,
which follows from just the counting of the Ramond fields.
The variation of $S^{(2)}_1$ under the gauge transformation $\delta_\Lambda V^{(0)}_0$
in (\ref{tf Lambda 0}) is calculated as
\begin{equation}
 (\delta_\Lambda)^{(0)}_0 S^{(2)}_1\ =\  \alpha_1\langle Q\Lambda,[\Psi^2]^\eta\rangle\
=\ -2\alpha_1\langle\Lambda,[\Psi,Q\Psi]^\eta\rangle\ =\
-2\alpha_1\langle[\Psi,\Lambda]^\eta,Q\Psi\rangle.
\end{equation}
This can be cancelled by $(\delta_\Lambda)^{(0)}_1 S^{(2)}_0$ if we take
\begin{equation}
 \delta_\Lambda\Psi^{(1)}_1\ =\ -2\alpha_1X\eta[\Psi,\Lambda]^\eta,
\end{equation}
in a similar manner given in \cite{Kunitomo:2015usa}.
Similarly, the variation of $S^{(2)}_1$ under the gauge transformation 
$\delta_\Omega V^{(0)}_0$ in (\ref{tf Omega 0}) is given by
\begin{equation}
 (\delta_\Omega)^{(0)}_0 S^{(2)}_1\ =\ \alpha_1\langle\eta\Omega, [\Psi^2]^\eta\rangle\ 
=\ -2\alpha_1 \langle\Omega,[\Psi, \eta\Psi]^\eta\rangle=\ 0,
\end{equation}
because of $\eta\Psi = 0$, and so we have
\begin{equation}
 \delta_\Omega\Psi^{(1)}_1\ =\ 0\,.
\label{omega1psi}
\end{equation}
Under the gauge transformation $\delta_\lambda\Psi^{(1)}_0$ 
in (\ref{tf lambda 0}), the variation of $S_1^{(2)}$ is given by
\begin{align}
 (\delta_\lambda)^{(0)}_0 S^{(2)}_1\ =&\ 2\alpha_1\langle Q\lambda,[V,\Psi]^\eta\rangle
\nonumber\\
=&\ 2\alpha_1\langle\lambda,[QV,\Psi]^\eta\rangle-2\alpha_1\langle\lambda,[V,Q\Psi]^\eta\rangle
\nonumber\\
=&\ 2\alpha_1\langle\Xi\lambda,[Q\eta V,\Psi]^\eta\rangle
+2\alpha_1\langle\Xi\lambda,[\eta V,Q\Psi]^\eta\rangle
\nonumber\\
=&\ 2\alpha_1\langle[\Psi,\Xi\lambda]^\eta,Q\eta V\rangle
+2\alpha_1\langle[\eta V,\Xi\lambda]^\eta,Q\Psi\rangle,
\label{l1}
\end{align}
where we used the fact that a relation,
\begin{equation}
\langle \lambda, B\rangle\ =\ \langle \eta\Xi\lambda, B\rangle
=\ \langle\Xi\lambda, \eta B\rangle,
\end{equation}
holds for general string field $B$ since the parameter $\lambda$ is in the small
Hilbert space.
This variation (\ref{l1}) can be canceled by 
$(\delta_\lambda)^{(2)}_1 S^{(0)}_0+(\delta_\lambda)^{(0)}_1 S^{(2)}_0$ 
with
\begin{equation}
 \delta_\lambda V^{(2)}_1\ =\ -2\alpha_1[\Psi,\Xi\lambda]^\eta,\qquad
 \delta_\lambda \Psi^{(1)}_1\ =\ 2\alpha_1 X\eta[\eta V,\Xi\lambda]^\eta.
\end{equation}

\subsubsection{Quartic interaction in $S_R$}

Let us move to the next order.
In order to narrow down the form of quartic interaction terms,
let us first consider the variation of $S^{(2)}_1$ under the gauge transformations
$\delta_\Lambda V^{(0)}_1$ and $\delta_\Lambda\Psi^{(1)}_1$, which is calculated as
\begin{align}
 (\delta_\Lambda)^{(0)}_1 S^{(2)}_1\ =&\
-\frac{\alpha_1}{2}\langle[V, Q\Lambda]^\eta,[\Psi^2]^\eta\rangle
-4\alpha_1^2\langle X\eta[\Psi,\Lambda]^\eta,[V,\Psi]^\eta\rangle
\nonumber\\
=&\ \frac{\alpha_1}{2}\langle Q\Lambda,[V,[\Psi^2]^\eta]^\eta\rangle
- 4\alpha_1^2\langle[\Psi,\Lambda]^\eta, X[\eta V,\Psi]^\eta\rangle\,.
\end{align}
Using $X=\{Q,\Xi\}$, we further calculate the second term 
$\langle[\Psi,\Lambda]^\eta, X[\eta V,\Psi]^\eta\rangle$ as follows:
\begin{align}
 \langle[\Psi,\Lambda]^\eta, X[\eta V,\Psi]^\eta\rangle\ =&\
\langle[\Psi,\Lambda]^\eta, \{Q,\Xi\}[\eta V,\Psi]^\eta\rangle
\nonumber\\
=&\ \langle[Q\Psi,\Lambda]^\eta,\Xi[\eta V,\Psi]^\eta\rangle
+\langle[\Psi,Q\Lambda]^\eta,\Xi[\eta V,\Psi]^\eta\rangle
\nonumber\\
&
- \langle\Xi [\Psi,\Lambda]^\eta, [Q\eta V,\Psi]^\eta\rangle
- \langle\Xi [\Psi,\Lambda]^\eta, [\eta V, Q\Psi]^\eta\rangle
\nonumber\\
=&\ -\langle [\Xi[\eta V,\Psi]^\eta,\Lambda]^\eta, Q\Psi\rangle
-\langle Q\Lambda,[\Psi,\Xi[\eta V,\Psi]^\eta]^\eta\rangle
\nonumber\\
& - \langle [\Psi,\Xi[\Psi,\Lambda]^\eta]^\eta,Q\eta V\rangle
- \langle[\eta V,\Xi[\Psi,\Lambda]^\eta]^\eta,Q\Psi\rangle\,.
\end{align}
Then we find
\begin{align}
  (\delta_\Lambda)^{(0)}_1 S^{(2)}_1\ =&\
\frac{\alpha_1}{2}\langle Q\Lambda,[V,[\Psi^2]^\eta]^\eta\rangle
+ 4\alpha_1^2 \langle Q\Lambda,[\Psi,\Xi[\eta V,\Psi]^\eta]^\eta\rangle
\nonumber\\
& + 4\alpha_1^2 \langle [\Psi,\Xi[\Psi,\Lambda]^\eta]^\eta,Q\eta V\rangle
\nonumber\\
&+ 4\alpha_1^2\langle [\Xi[\eta V,\Psi]^\eta,\Lambda]^\eta, Q\Psi\rangle
+ 4\alpha_1^2\langle[\eta V,\Xi[\Psi,\Lambda]^\eta]^\eta,Q\Psi\rangle\,.
\label{varLs2-1}
\end{align}
In order to cancel the first two terms on the right hand side, 
we introduce quartic interaction terms with two Ramond strings
as
\begin{equation}
 S^{(2)}_2\ =\ \alpha_2 \langle \Psi, [V, \eta V, \Psi]^\eta\rangle
+ \alpha_3 \langle \Psi, [V, \Xi[\eta V, \Psi]^\eta]^\eta\rangle,
\end{equation}
with constants $\alpha_2$ and $\alpha_3$ to be determined.
The first term is a genuine four-string interaction filling
a missing region in, for example, the moduli space of
four-string amplitude with two fermions.
The variation of $S^{(2)}_2$ under the gauge transformation
$(\delta_\Lambda)^{(0)}_0 V$ can be straightforwardly calculated as follows.
The variation of the first term 
is given by
\begin{align}
(\delta_\Lambda)^{(0)}_0(\alpha_2\langle\Psi,[V,\eta V,\Psi]^\eta\rangle)\ =&\
 \alpha_2\langle\Psi,[Q\Lambda, \eta V, \Psi]^\eta\rangle
+ \alpha_2\langle\Psi, [V, \eta Q\Lambda, \Psi]^\eta\rangle
\nonumber\\
=&\ \alpha_2\langle Q\Lambda, [\eta V, \Psi^2]^\eta\rangle
+\alpha_2\langle \eta Q\Lambda, [V, \Psi^2]^\eta\rangle
\nonumber\\
=&\ 2 \alpha_2\langle Q\Lambda, [\eta V, \Psi^2]^\eta\rangle
- \alpha_2\langle Q\Lambda, [V, [\Psi^2]^\eta]^\eta\rangle
\nonumber\\
&
+2 \alpha_2\langle Q\Lambda, [\Psi, [V, \Psi]^\eta]^\eta\rangle
\nonumber\\
=&\
- \alpha_2\langle Q\Lambda, [V, [\Psi^2]^\eta]^\eta\rangle
+2 \alpha_2\langle Q\Lambda, [\Psi, [V, \Psi]^\eta]^\eta\rangle
\nonumber\\
&\
- 2 \alpha_2\langle [\Psi^2,\Lambda]^\eta, Q\eta V\rangle
- 4 \alpha_2\langle [\eta V, \Psi, \Lambda]^\eta, Q\Psi\rangle\,.
\end{align}
The variation of the second term in $S^{(2)}_2$ can similarly be calculated as
\begin{align}
 (\delta_\Lambda)^{(0)}_0(\alpha_3\langle\Psi,[V, \Xi[\eta V,\Psi]^\eta]^\eta\rangle)\ =&\
\alpha_3\langle Q\Lambda, [\Psi,\Xi[\eta V,\Psi]^\eta]^\eta\rangle
+ \alpha_3 \langle \Psi, [V, \Xi[\eta Q\Lambda,\Psi]^\eta]^\eta\rangle,
\nonumber\\
=&\ \alpha_3\langle Q\Lambda,[\Psi,[V,\Psi]^\eta]^\eta\rangle
+ 2 \alpha_3\langle Q\Lambda,[\Psi,\Xi[\eta V, \Psi]^\eta]^\eta\rangle.
\end{align}
Therefore, in total, we have
\begin{align}
(\delta_\Lambda)^{(0)}_0 S^{(2)}_2\ =&\
- \alpha_2\langle Q\Lambda, [V, [\Psi^2]^\eta]^\eta\rangle
+ (2 \alpha_2+\alpha_3)\langle Q\Lambda, [\Psi, [V, \Psi]^\eta]^\eta\rangle
\nonumber\\
&
+2 \alpha_3\langle Q\Lambda,[\Psi,\Xi[\eta V, \Psi]^\eta]^\eta\rangle
\nonumber\\
&
- 2 \alpha_2\langle [\Psi^2,\Lambda]^\eta, Q\eta V\rangle
- 4 \alpha_2\langle [\eta V, \Psi, \Lambda]^\eta, Q\Psi\rangle.
\label{varLs2-2}
\end{align}
From (\ref{varLs2-1}) and (\ref{varLs2-2}), we find that
the constants $\alpha_1$, $\alpha_2$ and $\alpha_3$ should
be chosen to be
\begin{equation}
 \alpha_1\ =\ \frac{1}{2},\qquad \alpha_2\ =\ \frac{1}{4},\qquad
\alpha_3\ =\ -\,\frac{1}{2}\,.
\end{equation}
Then we have
\begin{align}
(\delta_\Lambda)^{(0)}_1 S^{(2)}_1\ + (\delta_\Lambda)^{(0)}_0 S^{(2)}_2\ =&\
%
- \frac{1}{2}\langle [\Psi^2,\Lambda]^\eta, Q\eta V\rangle
+ \langle [\Psi,\Xi[\Psi,\Lambda]^\eta]^\eta,Q\eta V\rangle
\nonumber\\
&
- \langle [\eta V, \Psi, \Lambda]^\eta, Q\Psi\rangle
+ \langle[\eta V,\Xi[\Psi,\Lambda]^\eta]^\eta,Q\Psi\rangle
\nonumber\\
&
+ \langle [\Xi[\eta V,\Psi]^\eta,\Lambda]^\eta, Q\Psi\rangle\,.
\end{align}
These terms can be cancelled by $(\delta_\Lambda)^{(2)}_2S^{(0)}_0$
and $(\delta_\Lambda)^{(0)}_2 S^{(2)}_0$ if we choose
\begin{align}
 \delta_\Lambda V^{(2)}_2\ =&\ \frac{1}{2}[\Psi^2,\Lambda]^\eta
- [\Psi,\Xi[\Psi,\Lambda]^\eta]^\eta,\\
 \delta_\Lambda \Psi^{(1)}_2\ =&\ - X\eta [\eta V, \Psi, \Lambda]^\eta
+ X\eta [\eta V, \Xi[\Psi, \Lambda]^\eta]^\eta
+ X\eta [\Xi[\eta V, \Psi]^\eta,\Lambda]^\eta.
\end{align}
Note that $\delta_\Lambda V^{(2)}_2 = (\delta_\Lambda)^{(2)}_2 V$
and $\delta_\Lambda \Psi^{(1)}_2 = (\delta_\Lambda)^{(0)}_2 \Psi$.
Thus
the gauge invariance under transformation with the parameter $\Lambda$ in this order
holds:
\begin{equation}
(\delta_\Lambda)^{(0)}_1 S^{(2)}_1\ + (\delta_\Lambda)^{(0)}_0 S^{(2)}_2
\ +(\delta_\Lambda)^{(2)}_2S^{(0)}_0 \ +(\delta_\Lambda)^{(0)}_2 S^{(2)}_0\ =\ 0\,.
\end{equation}

Then the variation under the gauge transformations with the parameter
$\Omega$ at this order can easily be calculated as
\begin{align}
 (\delta_\Omega)^{(0)}_1 S^{(2)}_1\ =&\
\frac{1}{4}\langle [\Psi^2]^\eta,[V, \eta\Omega]^\eta\rangle
=\
-\frac{1}{4}\langle\eta\Omega,[V,[\Psi^2]^\eta]^\eta\rangle
=\ -\frac{1}{4}\langle\Omega,[\eta V,[\Psi^2]^\eta]^\eta\rangle\,,
\\
  (\delta_\Omega)^{(0)}_0 S^{(2)}_2\ =&\
\frac{1}{4}\langle\eta\Omega,[\eta V,\Psi^2]^\eta\rangle
-\frac{1}{2}\langle\eta\Omega,[\Psi,\Xi[\eta V,\Psi]^\eta]^\eta\rangle
=\ \frac{1}{4}\langle\Omega,[\eta V,[\Psi^2]^\eta]^\eta\rangle\,,
\end{align}
and hence
\begin{equation}
 (\delta_\Omega)^{(0)}_1 S^{(2)}_1 + (\delta_\Omega)^{(0)}_0 S^{(2)}_2\ =\ 0\,.
\end{equation}
The correction at this order is not necessary:
\begin{equation}
 \delta_\Omega V^{(2)}_2\ =\ 0\,,\qquad
 \delta_\Omega \Psi^{(1)}_2\ =\ 0\,.
\end{equation}

Let us finally calculate variations of the action under the gauge transformation
with the parameter $\lambda$. The variations
$(\delta_\lambda)^{(2)}_1 S^{(0)}_1$ and $(\delta_\lambda)^{(0)}_1 S^{(2)}_1$ are
calculated as
\begin{align}
 (\delta_\lambda)^{(2)}_1 S^{(0)}_1\ =&\
- \frac{1}{2}\langle[QV, \eta V]^\eta, (\delta_\lambda)^{(2)}_1 V\rangle
=\ \frac{1}{2}\langle\Xi\lambda, [\Psi,[QV,\eta V]^\eta]^\eta\rangle\,,
\end{align}
and
\begin{align}
(\delta_\lambda)^{(0)}_1 S^{(2)}_1\ =&\
- \langle[V, \Psi]^\eta, X\eta [\eta V, \Xi\lambda]^\eta\rangle
=\ \langle \Xi\lambda, [\eta V, X [\eta V, \Psi]^\eta]^\eta\rangle,
\label{gt lambda 11}
\end{align}
respectively, where we used a relation 
\begin{alignat}{2}
 \langle A, X\eta B\rangle\ =&\  \langle \Xi\eta A, X\eta B\rangle\
=& \llangle \eta A, X\eta B\rrangle
\nonumber\\
=&\ \llangle X \eta A, \eta B\rrangle\
=& \langle\Xi X \eta A, \eta B\rangle
\nonumber\\
=&\ (-1)^A\langle X \eta A, B\rangle\,. &
\end{alignat}
%
The variation $(\delta_\lambda)^{(0)}_0 S^{(2)}_2$ 
is given by
\begin{align}
 (\delta_\lambda)^{(0)}_0 S^{(2)}_2\ 
=&\ 
\frac{1}{2}\langle Q\lambda, [V, \eta V, \Psi]^\eta\rangle
-\frac{1}{2}\langle Q\lambda, [V, \Xi[\eta V, \Psi]^\eta]^\eta\rangle
- \frac{1}{2}\langle Q\lambda, [\eta V,\Xi[V, \Psi]^\eta]^\eta\rangle\,.
\end{align}
Substituting the relation $Q\lambda=Q\eta\Xi\lambda$, this can further be
calculated as 
\begin{align}
(\delta_\lambda)^{(0)}_0 S^{(2)}_2\ =&\
- \frac{1}{2}\langle \Xi\lambda, [\Psi,[QV, \eta V]^\eta]^\eta\rangle
- \langle \Xi\lambda, [\eta V, X [\eta V,\Psi]^\eta]^\eta\rangle
\nonumber\\
&
+ \langle[\eta V, \Psi, \Xi\lambda]^\eta,Q\eta V\rangle
-\frac{1}{2}\langle [V, [\Psi,\Xi\lambda]^\eta]^\eta, Q\eta V\rangle
-\langle [\Xi[\eta V, \Psi]^\eta,\Xi\lambda]^\eta,Q\eta V\rangle
\nonumber\\
&
+\frac{1}{2}\langle [(\eta V)^2,\Xi\lambda]^\eta,Q\Psi\rangle
- \langle [\eta V, \Xi [\eta V, \Xi\lambda]^\eta]^\eta, Q\Psi\rangle
- \langle [\Psi, \Xi [\eta V, \Xi\lambda]^\eta]^\eta, Q\eta V\rangle
\nonumber\\
&
+\frac{1}{2}\langle [[V, \eta V]^\eta, \Xi\lambda]^\eta, Q\Psi\rangle.
\end{align}
In total we have
\begin{align}
(\delta_\lambda)^{(2)}_1 S^{(0)}_1+&(\delta_\lambda)^{(0)}_1 S^{(2)}_1
+ (\delta_\lambda)^{(0)}_0 S^{(2)}_2\ 
\nonumber\\
=&\ 
\langle[\eta V, \Psi, \Xi\lambda]^\eta,Q\eta V\rangle
-\frac{1}{2}\langle [V, [\Psi,\Xi\lambda]^\eta]^\eta, Q\eta V\rangle
- \langle [\Psi, \Xi [\eta V, \Xi\lambda]^\eta]^\eta, Q\eta V\rangle
\nonumber\\
&
-\langle [\Xi[\eta V, \Psi]^\eta,\Xi\lambda]^\eta,Q\eta V\rangle
+\frac{1}{2}\langle [(\eta V)^2,\Xi\lambda]^\eta,Q\Psi\rangle
- \langle [\eta V, \Xi [\eta V, \Xi\lambda]^\eta]^\eta, Q\Psi\rangle
\nonumber\\
&
+\frac{1}{2}\langle [[V, \eta V]^\eta, \Xi\lambda]^\eta, Q\Psi\rangle\,.
\end{align}
This can be cancelled by $(\delta_\lambda)^{(2)}_2 S^{(0)}_0$ and 
$(\delta_\lambda)^{(0)}_2 S^{(2)}_0$
if we take $(\delta_\lambda)^{(2)}_2 V$ and 
$(\delta_\lambda)^{(0)}_2 \Psi$ as
\begin{align}
 \delta_\lambda V^{(2)}_2\ =&\
-[\eta V, \Psi, \Xi\lambda]^\eta
+\frac{1}{2}[V,[\Psi,\Xi\lambda]^\eta]^\eta
+ [\Psi,\Xi[\eta V, \Xi\lambda]^\eta]^\eta
+ [\Xi[\eta V, \Psi]^\eta,\Xi\lambda]^\eta,\\
 \delta_\lambda \Psi^{(1)}_2\ =&\
\frac{1}{2}X\eta [(\eta V)^2,\Xi\lambda]^\eta
- X\eta [\eta V,\Xi[\eta V, \Xi\lambda]^\eta]^\eta
+\frac{1}{2} X\eta [[V,\eta V]^\eta,\Xi\lambda]^\eta.
\end{align}
So far so good:
In this way
the gauge invariance with the parameter $\lambda$ holds 
at quadratic order in both coupling constant and the Ramond fields:
\begin{equation}
(\delta_\lambda)^{(2)}_1 S^{(0)}_1 + (\delta_\lambda)^{(0)}_1 S^{(2)}_1
+ (\delta_\lambda)^{(0)}_0 S^{(2)}_2 + 
(\delta_\lambda)^{(2)}_2 S^{(0)}_0
+ (\delta_\lambda)^{(0)}_2 S^{(2)}_0 =0\,.
\end{equation}

Let us move to the quartic order in the Ramond string field.
The non-trivial contribution absent in the open superstring field theory
comes from $(\delta_\lambda)^{(2)}_1 S^{(2)}_1$ given by
\begin{align}
(\delta_\lambda)^{(2)}_1 S^{(2)}_1\ =&\ 
\frac{1}{2}\langle[\Psi^2]^\eta, \delta V_1^{(2)}\rangle\ 
=\ 
-\frac{1}{2}\langle \Xi\lambda,[\Psi,[\Psi^2]^\eta]^\eta\rangle\
=\
\frac{1}{6}\langle\lambda,[\Psi^3]^\eta\rangle,
\label{psi4}
\end{align}
which requires to add $\Psi^4$ interaction $S^{(4)}_2$ to the action.
Note that $\delta_\lambda \Psi_1^{(3)} ( = (\delta_\lambda)^{(2)}_1 \Psi )$ never appears. 
Let us consider (\ref{psi4}) in further detail.
As was given in (\ref{Q gauge product}),
$[\Psi^3]^\eta$ can be written   
in a \textit{BRST exact} form as,
\begin{equation}
[\Psi^3]^\eta\ =\ Q(\Psi^3)^\eta-3(\Psi^2,Q\Psi)^\eta,
\end{equation}
with the dual gauge product $(\cdots)^\eta$ given in (\ref{dual 3gp}).
Using this relation, (\ref{psi4}) can be rewritten as
\begin{align}
(\delta_\lambda)^{(2)}_1 S^{(2)}_1\ =&\ 
\frac{1}{6}\langle \lambda, Q(\Psi^3)^\eta\rangle
-\frac{1}{2}\langle\lambda,(\Psi^2,Q\Psi)^\eta\rangle 
\nonumber\\
=&\ 
-\frac{1}{6}\langle Q\lambda, (\Psi^3)^\eta\rangle
+ \frac{1}{2}\langle (\Psi^2,\lambda)^\eta, Q\Psi\rangle.
\end{align}
%
%
%
%
From the form of the first term, we can suppose that
$S^{(4)}_2$ have the form
\begin{equation}
 S^{(4)}_2\ =\ 
\alpha_4\langle\Psi, (\Psi^3)^\eta\rangle,
\end{equation}
with a constant $\alpha_4$, whose variation under $\delta_\lambda\Psi_0^{(1)}$
becomes
\begin{equation}
 (\delta_\lambda)^{(0)}_0 S^{(4)}_2\ 
=\ 4\alpha_4\langle Q\lambda,(\Psi^3)^\eta\rangle\,.
\end{equation}
Then we have
\begin{align}
(\delta_\lambda)^{(2)}_1 S^{(2)}_1 + (\delta_\lambda)^{(0)}_0 S^{(4)}_2\
=&\
\frac{1}{2} \langle(\Psi^2,\lambda)^\eta, Q\Psi\rangle\,,
\label{psi4-2}
\end{align}
by setting
\begin{equation}
\alpha_4\ =\ \frac{1}{4!}\,.
\end{equation}
The remaining terms
can be cancelled by $(\delta_\lambda)^{(2)}_2 S^{(2)}_0$
if we take
\begin{equation}
\delta_\lambda\Psi^{(3)}_2\ =\ 
\frac{1}{2}X\eta (\Psi^2,\lambda)^\eta.
\end{equation}
That is, the gauge invariance at this order holds:
\begin{equation}
(\delta_\lambda)^{(2)}_1 S^{(2)}_1 + (\delta_\lambda)^{(0)}_0 S^{(4)}_2
+(\delta_\lambda)^{(2)}_2 S^{(2)}_0 =0\,.
\end{equation}

Let us summarize the results up to this order. 
The action in the NS sector is given by
\begin{align}
S_{NS}\ =&\ S^{(0)}
\nonumber\\
 =&\ S^{(0)}_0 + \kappa S^{(0)}_1 + \kappa^2 S^{(0)}_2 + O(\kappa^3),
\end{align}
where
\begin{align}
 S^{(0)}_0\ =&\ -\frac{1}{2}\langle QV,\eta V\rangle,\\
 S^{(0)}_1\ =&\ -\frac{1}{3!}\langle QV, [V,\eta V]^\eta\rangle,\\
 S^{(0)}_2\ =&\ -\frac{1}{4!}\langle QV, [V,\eta V,\eta V]^\eta\rangle
-\frac{1}{4!}\langle QV,[V,[V,\eta V]^\eta]^\eta\rangle.
\end{align}
The action in the Ramond sector is given by
\begin{equation}
 S_{R}\ =\ S^{(2)}_0 + \kappa S^{(2)}_1 + \kappa^2 (S^{(2)}_2 + S^{(4)}_2) + O(\kappa^3),
\end{equation}
where
\begin{align}
 S^{(2)}_0\ =&\ -\frac{1}{2} \llangle\Psi, YQ\Psi\rrangle\,,\\
S^{(2)}_1\ =&\ \frac{1}{2}\langle\Psi,[V,\Psi]^\eta\rangle\,,\\
S^{(2)}_2\ =&\ \frac{1}{4} \langle\Psi,[V,\eta V,\Psi]^\eta\rangle
- \frac{1}{2} \langle\Psi,[V,\Xi[\eta V,\Psi]^\eta]^\eta\rangle\,,
\nonumber\\
S^{(4)}_2\ =&\ 
\frac{1}{4!}\langle\Psi,(\Psi^3)^\eta\rangle\,.
\end{align}
The gauge transformation with the gauge parameter for $\Lambda$ in the NS
sector is given by
\begin{align}
 \delta_\Lambda V\ =&\ \delta_\Lambda V^{(0)}_0 + \kappa \delta_\Lambda V^{(0)}_1
+ \kappa^2 \left(\delta_\Lambda V^{(0)}_2 + \delta_\Lambda V^{(2)}_2\right)  + O(\kappa^3)\,,\\
 \delta_\Lambda \Psi\ =&\ \delta_\Lambda \Psi^{(1)}_0 + \kappa \delta_\Lambda \Psi^{(1)}_1
+ \kappa^2 \delta_\Lambda \Psi^{(1)}_2 + O(\kappa^3)\,,
\end{align}
where
\begin{align}
 \delta_\Lambda V^{(0)}_0 \ =&\ Q\Lambda\,,\\
 \delta_\Lambda V^{(0)}_1\ =&\ -\frac{1}{2}[V,Q\Lambda]^\eta\,,\\
 \delta_\Lambda V^{(0)}_2\ =&\ -\frac{1}{3}[V,\eta V,Q\Lambda]^\eta
+\frac{1}{12}[V,[V,Q\Lambda]^\eta]^\eta\,,\\
\delta_\Lambda V^{(2)}_2\ =&\
 \frac{1}{2}[\Psi,\Psi,\Lambda]^\eta
- [\Psi,\Xi[\Psi,\Lambda]^\eta]^\eta\,,\\
\delta_\Lambda \Psi^{(1)}_0 \ =&\ 0\,,\\
\delta_\Lambda\Psi^{(1)}_1\ =&\ - X\eta [\Psi,\Lambda]^\eta\,,\\
\delta_\Lambda\Psi^{(1)}_2\ =&\ - X\eta[\eta V,\Psi,\Lambda]^\eta
+ X\eta [\eta V,\Xi[\Psi,\Lambda]^\eta]^\eta
+ X\eta [\Xi[\eta V, \Psi]^\eta,\Lambda]^\eta\,.
\end{align}
The gauge transformation with the gauge parameter $\Omega$ in the NS
sector is given by
\begin{align}
 \delta_\Omega V\ =&\ \delta_\Omega V^{(0)}_0 + \kappa \delta_\Omega V^{(0)}_1
+ \kappa^2 \delta_\Omega V^{(0)}_2 + O(\kappa^3)\,,\\
 \delta_\Omega \Psi\ =&\ \delta_\Omega \Psi^{(1)}_0 + \kappa \delta_\Omega \Psi^{(1)}_1
+ \kappa^2 \delta_\Omega \Psi^{(1)}_2 + O(\kappa^3)\,,
\end{align}
where
\begin{align}
 \delta_\Omega V^{(0)}_0\ =&\ \eta\Omega\,,\\
 \delta_\Omega V^{(0)}_1\ =&\ \frac{1}{2}[V,\eta\Omega]^\eta\,,\\
 \delta_\Omega V^{(0)}_2\ =&\ \frac{1}{6}[V,\eta V,\eta\Omega]^\eta
+ \frac{1}{12}[V,[V,\eta\Omega]^\eta]^\eta\,,\\
\delta_\Omega \Psi^{(1)}_0\ =&\ 0\,,\\
\delta_\Omega \Psi^{(1)}_1\ =&\ 0\,,\\
\delta_\Omega \Psi^{(1)}_2\ =&\ 0\,.
\end{align}
The gauge transformation with the gauge parameter $\lambda$ in the Ramond sector
is given by
\begin{align}
 \delta_\lambda V\ =&\ \delta_\lambda V^{(2)}_0 + \kappa \delta_\lambda V^{(2)}_1
+ \kappa^2 \delta_\lambda V^{(2)}_2 + O(\kappa^3)\,,\\
 \delta_\lambda \Psi\ =&\ \delta_\lambda \Psi^{(1)}_0 + \kappa \delta_\lambda \Psi^{(1)}_1
+ \kappa^2 \left(\delta_\lambda \Psi^{(1)}_2 + \delta_\lambda \Psi^{(3)}_2\right) + O(\kappa^3)\,,
\end{align}
where
\begin{align}
 \delta_\lambda V^{(2)}_0\ =&\ 0\,,\\
 \delta_\lambda V^{(2)}_1\ =&\ - [\Psi, \Xi\lambda]^\eta\,,\\
 \delta_\lambda V^{(2)}_2\ =&\ - [\eta V,\Psi, \Xi\lambda]^\eta
+\frac{1}{2}[V,[\Psi,\Xi\lambda]^\eta]^\eta
+ [\Psi,\Xi[\eta V,\Xi\lambda]^\eta]^\eta
+ [\Xi[\eta V,\Psi]^\eta,\Xi\lambda]^\eta\,,\\
 \delta_\lambda\Psi^{(1)}_0\ =&\ Q\lambda\,,\\
 \delta_\lambda\Psi^{(1)}_1\ =&\  X\eta [\eta V, \Xi\lambda]^\eta\,,\\
\delta_\lambda\Psi^{(1)}_2\ =&\ \frac{1}{2} X\eta [\eta V,\eta V,\Xi\lambda]^\eta
- X\eta [\eta V,\Xi[\eta V,\Xi\lambda]^\eta]^\eta
+\frac{1}{2} X\eta [[V,\eta V]^\eta, \Xi\lambda]^\eta\,,\\
\delta_\lambda \Psi^{(3)}_2\ =&\ \frac{1}{2}X\eta 
(\Psi^2,\lambda)^\eta\,.
\end{align}

\subsection{Fermion expansion}

As a next step to the complete action let us consider the fermion expansion.
We extend the above results to all order in the NS string field at each order in the
Ramond string field.

Suppose that an arbitrary variation of $S^{(2n)}$, the action at $O(\Psi^{2n})$, has the form
\begin{align}
\delta S^{(2n)} = -\langle \!\langle \delta \Psi, Y E^{(2n-1)}\rangle \!\rangle 
+ \langle B_\delta , E^{(2n)}\rangle\,.
\label{var}
\end{align}
The equations of motion are therefore given by
\begin{align}
 E^{(0)}+E^{(2)}+E^{(4)}+\cdots\ =\ 0\,,\\
 E^{(1)}+E^{(3)}+E^{(5)}+\cdots\ =\ 0\,,
\end{align}
for the NS and the Ramond string fields, respectively.
We can also expand the gauge transformation in powers of the Ramond string field as
\begin{align}
B_\delta &= B_\delta^{(0)} +B_\delta^{(2)} +B_\delta^{(4)} +\cdots,\\
\delta \Psi & =\delta \Psi^{(1)} +\delta \Psi^{(3)} + \delta\Psi^{(5)} +\cdots.
\end{align}
Note that the superscript also counts the Ramond gauge parameter $\lambda$.
We will determine the action and the gauge transformations order by order 
in the Ramond string field by
requiring the gauge-invariance at each order:
\begin{align}
0= -\sum^{n}_{k=1} \langle \!\langle \delta \Psi ^{(2n-2k+1)}, Y E^{(2k-1)}\rangle \!\rangle 
+ \sum^{n}_{k=0}\langle B_\delta^{(2n-2k)} , E^{(2k)}\rangle\,.
\label{fermion exp gen}
\end{align}
In particular, at the lowest order in fermion, corresponding to $n=0$,
it reduces 
\begin{equation}
 0\ =\ \langle B_\delta^{(0)},E^{(0)}\rangle.
\label{gen eom 0}
\end{equation}
For the dual WZW-like action (\ref{dual action NS}), $E^{(0)}=QG_\eta(V)$, 
and (\ref{gen eom 0}) requires  
$B_\delta^{(0)}=Q\Lambda+D_\eta\Omega$ as summarized in the previous section.

\subsubsection{Quadratic in fermion}

We first provide the action $S^{(2)}$ and the gauge transformation $B_\delta^{(2)}$ and $\delta \Psi^{(1)}$
so that the action is gauge invariant at quadratic order in the Ramond string field:
\begin{align}
0=-\langle \!\langle \delta \Psi ^{(1)}, Y E^{(1)}\rangle \!\rangle +\langle B_\delta^{(2)} , 
E^{(0)}\rangle+\langle B_\delta^{(0)} , E^{(2)}\rangle.
\label{fermion exp quadratic}
\end{align}
From the results in the perturbative expansion, we can deduce that the action $S^{(2)}$
is given by the same form of that for the open superstring (\ref{comp action open}): 
\begin{align}
S^{(2)}&=-\frac{1}{2}\langle\!\langle \Psi,  YQ \Psi\rangle\!\rangle 
+\frac{\kappa}{2}\int^1_0 dt  \big\langle B_t(t),  [F(t)\Psi,F(t)\Psi]^\eta_{G_\eta(t)}\big\rangle,
\label{quadratic action}
\end{align}
where $F(t)$ is the linear operator defined by
\begin{align}
F(t)=\frac{1}{1+\Xi(D_\eta(t) -\eta)}
=1 + \sum_{n=1}^\infty \big(-\Xi(D_\eta(t)-\eta)\big)^n.
\label{F hetero}
\end{align}
We should note that this has the same form as (\ref{F open}) but
$D_\eta$ defined in (\ref{shifted L-infty}) 
contains infinite number of terms with arbitrary power of $G_\eta$.
In what follows, we show that the variation of the action $S^{(2)}$ becomes
\begin{align}
\delta S^{(2)} 
&= -\langle\!\langle \delta\Psi,  Y( E^{(1)}) \rangle\!\rangle
+ \big\langle B_\delta,E^{(2)}\big\rangle
\nonumber\\
&=-\langle\!\langle \delta \Psi, Y(Q\Psi+X\eta F\Psi)\rangle\!\rangle 
+ \frac{\kappa}{2} \big\langle B_\delta , [F\Psi,F\Psi]^\eta_{G_\eta}\big\rangle\,,
\end{align}
and prove that the action $S^{(0)}+S^{(2)}$ is invariant at quadratic order in the Ramond string field
under the following gauge transformations at this order
\begin{align}
%
%
%
B_{\delta}^{(2)}\ =&\
 \frac{\kappa^2}{2} [F\Psi,F\Psi,\Lambda]^\eta_{G_\eta}
-\kappa^2[F\Psi, F\Xi[F\Psi,\Lambda]^\eta_{{G_\eta}}]^\eta_{{G_\eta}}
-\kappa[ F\Psi,F\Xi \lambda]^\eta_{G_\eta}\,,\\
\delta\Psi^{(1)}\ =&\ -\kappa X\eta  F\Xi D_\eta [F\Psi,\Lambda]^\eta_{G_\eta} 
+Q\lambda + X\eta F\lambda\,.
\label{quadratic gauge tf}
\end{align}

Let us first summarize the properties of 
$F(t)$, we will use.
The linear map $F(t)$ satisfies $F(0) = 1$ since it depends on $t$ only through $G_\eta(t)$ 
and $G_\eta(0)=0$. It is invertible and 
\begin{equation}
F^{-1}(t)=1+\Xi(D_\eta(t) -\eta)=\eta \Xi +\Xi D_\eta(t)\,. 
\end{equation}
Multiplying it by $\eta$ from the left or by $D_\eta$ from the right, we have
\begin{equation}
 \eta F^{-1}(t)\ =\ F^{-1}(t)D_\eta(t)=\ \eta\Xi D_\eta(t)\,,
\label{eta Finverse}
\end{equation}
We further have
\begin{equation}
F(t)\eta\ =\ D_\eta(t) F(t)\,,\qquad
\{D_\eta(t), F(t)\Xi\}\ =\ 1\,.
\end{equation}
The former can be obtained by multiplying the first equation in (\ref{eta Finverse})
by F(t) from both left and right and using $\eta^2=D_\eta(t)^2=0$.
Then the latter can be derived as
\begin{align}
D_\eta(t) F(t)\Xi + F(t)\Xi D_\eta(t) 
&= F(t)\eta \Xi +F(t)\Xi D_\eta (t)\no
&=F(t) \big( 1+\Xi(D_\eta(t) -\eta) \big)\ =\ 1\,.
\end{align}
It is also shown
\begin{equation}
 \langle F(t)\Xi V_1, V_2\rangle\ =\ (-1)^{V_1+1} \langle  V_1 , F(t)\Xi V_2\rangle\,,
\end{equation}
from the definition (\ref{F hetero}) and the BPZ properties (\ref{partial Qeta}),
 (\ref{partial Deta}), and (\ref{BPZ Xi}).
The commutator of $F(t)$ and the derivation 
$\mathbb{X}=Q$, $\partial_t$, or $\delta$
on the dual string products is given by
\begin{align}
[\mathbb X, F(t) ] V_1\
=&\ - F(t) [\mathbb X, F^{-1}(t)] F(t) V_1\no
=&\ -F(t) (\mathbb{X}\Xi-(-1)^{\mathbb{X}}\Xi\mathbb{X}) (D_\eta(t) -\eta) F(t)V_1 
\nonumber\\
&\
-\kappa F(t)\Xi [\mathbb X {G_\eta(t)}, F(t)V_1 ]^\eta_{{G_\eta(t)}}\,.\label{FXCOMMU}
\end{align}
%
We also summarize the properties of $F(t)\Psi$ for later use. 
Since $F(t)\eta = D_\eta(t) F(t)$ and $\eta \Psi=0$, $F(t)\Psi$ is $D_\eta(t)$-exact:
\begin{align}
F(t)\Psi = F(t) \{\eta, \Xi \}\Psi = D_\eta(t) F(t)\Xi \Psi.
\end{align}
Acting with $QF(t)$ on $\Psi$, (\ref{FXCOMMU}) leads to
\begin{align}
QF(t)\Psi\ =&\ F(t) \big( Q \Psi + X \eta F(t)\Psi \big) 
- \kappa F(t)\Xi [ Q G_\eta(t), F(t)\Psi]^\eta_{G_\eta(t)}
\nonumber\\
=&\ D_\eta(t)F\Xi(Q\Psi+X\eta F(t)\Psi)
-\kappa F(t)\Xi[QG_\eta(t),F(t)\Psi]^\eta_{G_\eta(t)}.
\label{QFPsi}
\end{align}
For $\mathbb{X}=\partial_t$, or $\delta$,
which commute with $\Xi$, 
$\mathbb{X}F(t)\Psi$
can be transformed into the following form:
\begin{align}
\mathbb X F(t)\Psi\ 
=&\ F(t) \mathbb X \Psi 
+ (-1)^\mathbb X\kappa F(t) \Xi D_\eta(t) [B_\mathbb X(t),F(t)\Psi]^\eta_{{G_\eta(t)}}\label{XF COM1}\\
=&\ F(t) \mathbb X \Psi + (-1)^\mathbb X \kappa [B_\mathbb X(t),F(t)\Psi]^\eta_{{G_\eta(t)}}
\nonumber\\
&\
- (-1)^\mathbb X \kappa D_\eta(t) F(t) \Xi [B_\mathbb X(t),F(t)\Psi]^\eta_{{G_\eta(t)}},
\label{XF COM3}
\end{align}
where we used
 $\mathbb X {G_\eta(t)} =(-1)^\mathbb X D_\eta(t) B_\mathbb X(t)$ 
and $\{ D_\eta(t) , F(t)\Xi \}=1$.
%
%

Now let us consider the variation of $S^{(2)}$: 
\begin{equation}
\delta S^{(2)}\ =\ -\llangle \delta\Psi, YQ\Psi\rrangle
+\frac{\kappa}{2} \int^1_0 dt\ \delta\langle B_t(t),[F(t)\Psi,F(t)\Psi]^\eta_{G_\eta(t)}\rangle.
\end{equation}
From here to (\ref{869}), we omit the $t$-dependence for notational brevity.
The variation of the integrand of the interaction term,
\begin{align}
\frac{\kappa}{2}\delta \big\langle B_t, [F\Psi,F\Psi]^\eta_{{G_\eta}}\big\rangle\
=&\ \frac{\kappa}{2} \big\langle \delta B_t, [F\Psi,F\Psi]^\eta_{{G_\eta}}\big\rangle
+\kappa\big\langle B_t,  [\delta F\Psi,F\Psi]^\eta_{{G_\eta}}\big\rangle
\nonumber\\
&\
+\frac{\kappa^2}{2} \big\langle B_t, [\delta {G_\eta}, F\Psi,F\Psi]^\eta_{{G_\eta}}\big\rangle,
\label{VAR INTEGRAND}
\end{align}
can be calculated as follows.
Since $[F\Psi,F\Psi]^\eta_{{G_\eta}}$ is $D_\eta$-exact,
we can use (\ref{dual FXY}) for the first term, and obtain
\begin{align}
\frac{\kappa}{2} \big\langle \delta B_t, [F\Psi,F\Psi]^\eta_{{G_\eta}}\big\rangle\
=&\ \frac{\kappa}{2} \big\langle \partial_t B_\delta +\kappa[B_\delta, B_t]^\eta_{{G_\eta}} , 
[F\Psi,F\Psi]^\eta_{{G_\eta}}\big\rangle
\nonumber\\
=&\
\frac{\kappa}{2}\big\langle \partial_t B_\delta, [F\Psi, F\Psi]^\eta_{G_\eta}\big\rangle
+\frac{\kappa^2}{2}\big\langle B_\delta, 
[B_t, [F\Psi, F\Psi]^\eta_{G_\eta}]^\eta_{G_\eta}\big\rangle.\label{VOI 1}
\end{align}
For the second term, 
utilizing (\ref{XF COM3}), we find
\begin{align}
\kappa\big\langle B_t,  [\delta F\Psi,F\Psi]^\eta_{{G_\eta}}\big\rangle\
&=\ -\kappa\big\langle [B_t,F\Psi]^\eta_{{G_\eta}},  F\delta\Psi + \kappa [B_\delta, F\Psi]^\eta_{{G_\eta}} 
-\kappa D_\eta F \Xi [B_\delta, F\Psi]^\eta_{{G_\eta}} \big\rangle\no
&=\ -\kappa\big\langle [B_t,F\Psi]^\eta_{{G_\eta}},   
\kappa [B_\delta, F\Psi]^\eta_{{G_\eta}} + D_\eta F\Xi\left( \delta\Psi
-\kappa [B_\delta, F\Psi]^\eta_{{G_\eta}}\right) \big\rangle\no
&=\ -\kappa^2 \big\langle [B_t,F\Psi]^\eta_{{G_\eta}},  [B_\delta, F\Psi]^\eta_{{G_\eta}} \big\rangle
-\big\langle \partial_t F\Psi, \delta\Psi- \kappa [B_\delta, F\Psi]^\eta_{{G_\eta}} \big\rangle\no
&=\ -\kappa^2 
\big\langle B_\delta, [F\Psi, [B_t, F\Psi]^\eta_{G_\eta}]^\eta_{G_\eta}\big\rangle
+\big\langle  \delta \Psi , \partial_t F\Psi\big\rangle
+\kappa\big\langle B_\delta,  [ \partial_t F \Psi, F\Psi]^\eta_{{G_\eta}} \big\rangle.\label{VOI 2}
\end{align}
For the third term, 
utilizing the $L_\infty$-relation of 
${G_\eta}$-shifted dual string products, we obtain
\begin{align}
\frac{\kappa^2}{2} \big\langle B_t, [\delta {G_\eta}, F\Psi,F\Psi]^\eta_{{G_\eta}}\big\rangle\
=&\ \frac{\kappa^2}{2} \big\langle B_t, [D_\eta B_\delta, F\Psi,F\Psi]^\eta_{{G_\eta}}\big\rangle\no
=&\ \frac{\kappa^2}{2} \big\langle B_t, \Big(- D_\eta[ B_\delta, F\Psi,F\Psi]^\eta_{{G_\eta}}
+ [B_\delta, [F\Psi,F\Psi]^\eta_{{G_\eta}}]^\eta_{{G_\eta}}
\nonumber\\
&\hspace{20mm}
-2[F\Psi, [B_\delta, F\Psi]^\eta_{{G_\eta}}]^\eta_{{G_\eta}}
\Big)\big\rangle\no
=&\
\frac{\kappa^2}{2} \big\langle B_\delta, [ \partial_t {G_\eta}, F\Psi,F\Psi]^\eta_{{G_\eta}}\big\rangle
-\frac{\kappa^2}{2}\big\langle B_\delta, [B_t, [F\Psi, F\Psi]^\eta_{G_\eta}]^\eta_{G_\eta}\big\rangle
\nonumber\\
&
+ \kappa^2 \big\langle B_\delta, [F\Psi, [B_t, F\Psi]^\eta_{G_\eta}]^\eta_{G_\eta}\big\rangle\,.
\label{VOI 3}
\end{align}
Then the total variation is given by
\begin{align}
&\frac{\kappa}{2} \delta \big\langle B_t, [F\Psi,F\Psi]^\eta_{{G_\eta}}\big\rangle\no
&=
\frac{\kappa}{2} \big\langle \partial_t B_\delta , [F\Psi,F\Psi]^\eta_{{G_\eta}}\big\rangle
+\frac{\kappa^2}{2} \big\langle B_\delta, [ \partial_t {G_\eta}, F\Psi,F\Psi]^\eta_{{G_\eta}}\big\rangle
+\kappa\big\langle B_\delta,  [\partial_t F \Psi, F\Psi]^\eta_{{G_\eta}} \big\rangle
+\big\langle  \delta \Psi , \partial_t F\Psi\big\rangle\no
&=
\partial_t \Big( 
- \big\llangle  \delta \Psi , YX\eta F\Psi\big\rrangle +
\frac{\kappa}{2} \big\langle B_\delta , [F\Psi,F\Psi]^\eta_{{G_\eta}}\big\rangle\Big)\,,
\label{869}
\end{align}
where we assumed that $\delta\Psi$ also satisfies the constraints (\ref{constR1}).
%
%
Using $B_\delta(0)=0$ and $\eta F(0)\Psi=\eta\Psi=0$, we eventually find
\begin{align}
 \delta S^{(2)}\ 
=&\
-\llangle\delta\Psi, Y(Q\Psi + X\eta F\Psi)\rrangle
+ \frac{\kappa}{2}\langle B_\delta,[F\Psi,F\Psi]^\eta_{G_\eta}\rangle\,,
\end{align}
and hence
\begin{equation}
 E^{(1)}\ =\ Q\Psi+X\eta F\Psi\,,\qquad
 E^{(2)}\ =\ \frac{\kappa}{2}[F\Psi,F\Psi]^\eta_{G_\eta}\,.
\label{E one and two}
\end{equation}

By requiring (\ref{fermion exp quadratic}),
let us determine $\delta\Psi^{(1)}$ and $B_\delta^{(2)}$
for each of gauge transformations with the parameters $\Lambda$, $\Omega$ and $\lambda$.
%
%
%
%
Let us first consider the invariance under the transformation with the parameter $\Lambda$:
\begin{align}
0\ =&\ -\langle\!\langle {\delta_\Lambda}\Psi^{(1)},  YE^{(1)} \rangle\!\rangle
+ \big\langle B_{\delta_\Lambda}^{(0)} ,E^{(2)}\big\rangle 
+ \big\langle B_{\delta_\Lambda}^{(2)} ,E^{(0)}\big\rangle\,.
\label{GI R2 L}
\end{align}
Here the second term is already known. Recalling (\ref{QFPsi}) (at $t=1$),
\begin{equation}
 QF\Psi\ =\ D_\eta F\Xi E^{(1)}-\kappa F\Xi[E^{(0)},F\Psi]^\eta_{G_\eta},
\end{equation}
it can be calculated as
\begin{align}
\big\langle B_{\delta_\Lambda}^{(0)} ,E^{(2)}\big\rangle
=&\ \frac{\kappa}{2}\big\langle Q\Lambda , [F\Psi,F\Psi]^\eta_{G_\eta}\big\rangle
\nonumber\\
=&\
\langle\Big(
-\frac{\kappa^2}{2}[F\Psi,F\Psi,\Lambda]^\eta_{G_\eta}
+\kappa^2[F\Psi,F\Xi[F\Psi,\Lambda]^\eta_{G_\eta}]^\eta_{G_\eta}\Big),E^{(0)}\rangle
\nonumber\\
&
-\kappa\langle F\Xi D_\eta[F\Psi,\Lambda]^\eta_{G_\eta},E^{(1)}\rangle\,.
\label{GT R2 Lam1}
\end{align}
If we note that $E^{(1)}$ satisfies the constraints (\ref{constR1}),
(\ref{GI R2 L}) holds by taking
\begin{align}
B_{\delta_\Lambda}^{(2)}\ =&\ 
\frac{\kappa^2}{2}[F\Psi,F\Psi,\Lambda]^\eta_{G_\eta}
-\kappa^2[F\Psi,F\Xi[F\Psi,\Lambda]^\eta_{G_\eta}]^\eta_{G_\eta}\,,\\
{\delta_\Lambda}\Psi^{(1)}\ =&\ 
- \kappa X\eta F\Xi D_\eta[F\Psi,\Lambda]^\eta_{G_\eta}\,.
 \end{align}
The invariance under the transformation with the parameter $\Omega$ requires
\begin{align}
0= -\langle\!\langle {\delta_\Omega}\Psi^{(1)},  YE^{(1)} \rangle\!\rangle
+ \big\langle B_{\delta_\Omega}^{(0)} ,E^{(2)}\big\rangle + \big\langle B_{\delta_\Omega}^{(2)} ,E^{(0)}\big\rangle\,.
\label{GT R2 O}
\end{align}
Since the second term is again known and calculated as
\begin{equation}
\langle B_{\delta_\Omega}^{(0)} ,E^{(2)}\rangle\ =\ \langle D_\eta\Omega,E^{(2)}\rangle\
=\ \langle\Omega, D_\eta E^{(2)}\rangle\ =\ 0\,,
\end{equation}
%
we conclude that
\begin{align}
B_{\delta_\Omega}^{(2)} =0,\qquad
\delta_\Omega\Psi^{(1)} =0\,.
\end{align}
%
%
Finally, 
for
the invariance under the transformation with $\lambda$:
\begin{align}
0\ =&\ -\langle\!\langle {\delta_\lambda}\Psi^{(1)},  YE^{(1)} \rangle\!\rangle
+ \big\langle B_{\delta_\lambda}^{(2)} ,E^{(0)}\big\rangle
\nonumber\\
=&\ -\llangle\delta_\lambda\Psi_0^{(1)}, YE^{(1)}\rrangle
-\llangle \tilde{\delta}_\lambda\Psi^{(1)},YE^{(1)}\rrangle
+ \big\langle B_{\delta_\lambda}^{(2)} ,E^{(0)}\big\rangle\,,
\label{GI R2 l}
\end{align}
where we decomposed $\delta_\lambda\Psi^{(1)}$ into the free part (\ref{tf lambda 0})
and remaining: $\delta_\lambda \Psi^{(1)} =\delta_\lambda \Psi^{(1)}_0 + \tilde{\delta}_\lambda \Psi^{(1)}$.
The known part in this case is the first term, which is calculated as
\begin{align}
 -\llangle\delta_\lambda\Psi_0^{(1)}, YE^{(1)}\rrangle\
=&\ -\llangle Q\lambda, Y(Q\Psi+X\eta F\Psi)\rrangle\
 =\ 
\big\langle Q\lambda, F\Psi \big\rangle\
\nonumber\\
=&\
\kappa\langle[F\Psi, F\Xi\lambda]^\eta_{G_\eta},E^{(0)}\rangle
-\langle F\Xi D_\eta\lambda, E^{(1)}\rangle\,.
\label{GT R2 lam 1}
\end{align}
The invariance (\ref{GI R2 l}) holds if we take
\begin{align}
B_{\delta_{\lambda}}^{(2)}\ =&\ -\kappa[ F\Psi ,F\Xi \lambda]^\eta_{G_\eta},\\
\tilde{\delta}_{\lambda}\Psi^{(1)}\ =&\ - X\eta F\Xi D_\eta\lambda\
=\ X\eta F\lambda\,.\label{tf lambda}
\end{align}
Thus, in total, the gauge transformation at this order becomes (\ref{quadratic gauge tf}).

\subsubsection{Quartic in fermion}

So far we have determined the complete action at the quadratic order in fermion
(\ref{quadratic action}), which has the same form as that of the open superstring
field theory, and thus is its straightforward extension.
For the heterotic string field theory, however, this is not the end of story.
At the next order in the fermion expansion, the gauge invariance further requires
%
\begin{align}
0
=-\big\langle\!\big\langle \delta\Psi^{(1)}, YE^{(3)}\big\rangle\!\big\rangle 
-\big\langle\!\big\langle \delta \Psi^{(3)}, YE^{(1)}\big\rangle\!\big\rangle
+ \big\langle B_{\delta}^{(0)}, E^{(4)}\big\rangle
+ \big\langle B_{\delta}^{(2)}, E^{(2)}\big\rangle
+ \big\langle B_{\delta}^{(4)}, E^{(0)}\big\rangle\,,\label{GI R4}
\end{align}
%
%
in which, in particular, 
we find
%
%
%
\begin{equation}
 \langle B_\delta^{(2)}, E^{(2)}\rangle\ \ne 0\,.
\end{equation}
Thus it is necessary to add the action $S^{(4)}$ quartic in fermion, 
and determine $B_\delta^{(4)}$ and $\delta\Psi^{(3)}$
so that the equation (\ref{GI R4}) is satisfied.
%

%
%
%
%

Let us begin with considering the transformation with the parameter $\lambda$,
which is 
the most efficient 
way to find out $S^{(4)}$ as shown in the following. 
From (\ref{E one and two}) and (\ref{tf lambda}) we have
\begin{align}
 \langle B_{\delta_\lambda}^{(2)},E^{(2)}\rangle\ 
=&\ \frac{\kappa^2}{6}\langle F\Xi\lambda, D_\eta[F\Psi,F\Psi,F\Psi]^\eta_{G_\eta}\rangle
\nonumber\\
=&\ \frac{\kappa^2}{6}\langle F\lambda,[F\Psi,F\Psi,F\Psi]^\eta_{G_\eta}\rangle\,.
\label{B2E2}
\end{align}
Here, from the $Q$-exactness of the dual string products (\ref{SDGP}), we can rewrite
\begin{align}
 [F\Psi,F\Psi,F\Psi]^\eta_{G_\eta}\ =&\ Q(F\Psi,F\Psi,F\Psi)^\eta_{G_\eta}
-3(F\Psi,F\Psi,QF\Psi)^\eta_{G_\eta}
\nonumber\\
&
-\kappa (QG_\eta,F\Psi,F\Psi,F\Psi)^\eta_{G_\eta}\,,
\end{align}
and thus
\begin{align}
\langle B_{\delta_\lambda}^{(2)},E^{(2)}\rangle\
=&\ 
\frac{\kappa^2}{6}\langle F\lambda, Q(F\Psi,F\Psi,F\Psi)^\eta_{G_\eta}\rangle
-\frac{\kappa^2}{2}\langle F\lambda, (F\Psi,F\Psi,QF\Psi)^\eta_{G_\eta}\rangle
\nonumber\\
&
-\frac{\kappa^3}{6}\langle F\lambda, (QG_\eta,F\Psi,F\Psi,F\Psi)^\eta_{G_\eta}\rangle
\nonumber\\
=&\
-\frac{\kappa^2}{6}\langle QF\lambda, (F\Psi,F\Psi,F\Psi)^\eta_{G_\eta}\rangle\
+ \frac{\kappa^2}{2}\langle (F\Psi,F\Psi,F\lambda)^\eta_{G_\eta},QF\Psi\rangle\
\nonumber\\
&
+\frac{\kappa^3}{6}\langle (F\Psi,F\Psi,F\Psi,F\lambda)^\eta_{G_\eta},QG_\eta\rangle\,.
\end{align}
Using (\ref{FXCOMMU}), the first and second terms can further be calculated as
\begin{align}
-\frac{\kappa^2}{6}\langle QF\lambda, (F\Psi,F\Psi,F\Psi)^\eta_{G_\eta}\rangle\
=&\
\frac{\kappa^2}{6}\llangle (Q\lambda+X\eta F\lambda), 
YX\eta F\Xi D_\eta(F\Psi,F\Psi,F\Psi)^\eta_{G_\eta}\rrangle
\nonumber\\
&
-\frac{\kappa^3}{6}\langle [F\Xi(F\Psi,F\Psi,F\Psi)^\eta_{G_\eta},F\lambda]^\eta_{G_\eta},
QG_\eta\rangle\,,
\end{align}
and
\begin{align}
\frac{\kappa^2}{2}\langle (F\Psi,F\Psi,F\lambda)^\eta_{G_\eta},QF\Psi\rangle\
=&\
\frac{\kappa^2}{2}\llangle X\eta F\Xi D_\eta(F\Psi,F\Psi,F\lambda)^\eta_{G_\eta},
Y(Q\Psi+X\eta F\Psi)\rrangle
\nonumber\\
&
-\frac{\kappa^3}{2}\langle[F\Psi,F\Xi(F\Psi,F\Psi,F\lambda)^\eta_{G_\eta}]^\eta_{G_\eta},
QG_\eta\rangle\,,
\end{align}
respectively, and we eventually have
\begin{align}
   \langle B_{\delta_\lambda}^{(2)},E^{(2)}\rangle\ =&\
\frac{\kappa^2}{6}\llangle\delta_\lambda\Psi^{(1)},Y
X\eta F\Xi D_\eta(F\Psi,F\Psi,F\Psi)^\eta_{G_\eta}\rrangle
\nonumber\\
&
+\frac{\kappa^2}{2}\llangle X\eta F\Xi D_\eta(F\Psi,F\Psi,F\lambda)^\eta_{G_\eta}, YE^{(1)}\rrangle
\nonumber\\
&
+\langle\Big(\frac{\kappa^3}{6}(F\Psi,F\Psi,F\Psi,F\lambda)^\eta_{G_\eta}
-\frac{\kappa^3}{2}[F\Psi,F\Xi(F\Psi,F\Psi,F\lambda)^\eta_{G_\eta}]^\eta_{G_\eta}
\nonumber\\
&\hspace{20mm}
-\frac{\kappa^3}{6}[F\Xi(F\Psi,F\Psi,F\Psi)^\eta_{G_\eta},F\lambda]^\eta_{G_\eta}
\Big), E^{(0)}\rangle\,.
\end{align}
Substituting this into (\ref{GI R4}), and taking into account
$B_{\delta_\lambda}^{(0)}=0$, we obtain
\begin{align}
 E^{(3)}\ =&\ \frac{\kappa^2}{6} X \eta F\Xi D_\eta (F\Psi,F\Psi,F\Psi)^\eta_{G_\eta}\,,\\
B_{\delta_\lambda}^{(4)}\ 
=&\ 
-\frac{\kappa^3}{6} (F\Psi,F\Psi,F\Psi,F\lambda)^\eta_{G_\eta}
+\frac{\kappa^3}{6} [F\Xi (F\Psi,F\Psi,F\Psi)^\eta_{G_\eta},F \lambda]^\eta_{{G_\eta}} 
\nonumber\\
&
+\frac{\kappa^3}{2} [F \Psi, F\Xi (F\Psi,F\Psi,F\lambda)^\eta_{G_\eta}]^\eta_{{G_\eta}}\,,\\
\delta_\lambda\Psi^{(3)}\ =&\ \frac{\kappa^2}{2} X\eta F \Xi D_\eta (F\Psi,F\Psi,F\lambda)^\eta_{G_\eta}\,.
\end{align}
%
%
%
From this form of $E^{(3)}$, the action
$S^{(4)}$ has to satisfy
\begin{align}
 \delta S^{(4)}\ =&\ -\llangle \delta\Psi, Y E^{(3)}\rrangle
\nonumber\\
=&\
-\frac{\kappa^2}{6} \big\langle\!\big\langle  \delta \Psi, 
Y X\eta F\Xi D_\eta(F\Psi,F\Psi,F\Psi)^\eta_{G_\eta}\big\rangle\!\big\rangle\no
=&\ \frac{\kappa^2}{6} \big\langle  F \delta \Psi  , (F\Psi,F\Psi,F\Psi)^\eta_{G_\eta}\big\rangle\,,
\label{933}
\end{align}
under an arbitrary variation of $\Psi$, where we used $\delta\Psi$ satisfies the constraint 
(\ref{constR1}) and therefore $D_\eta F\Xi \delta \Psi = F\eta\Xi\delta\Psi = F \delta \Psi$.
Since the shifted dual gauge products are cyclic, we can integrate it, and obtain
\begin{align}
 S^{(4)} &=\frac{\kappa^2}{24} \big\langle  F \Psi  ,  (F\Psi,F\Psi,F\Psi)^\eta_{G_\eta}\big\rangle\,.
\label{quartic action}
\end{align}

We further consider the gauge transformations in the NS sector. Under an arbitrary
variation of the NS string field, we have
\begin{align}
\delta S^{(4)} 
=&\ \frac{\kappa^3}{6} \big\langle  -F\Xi[\delta {G_\eta}, F\Psi]^\eta_{{G_\eta}}  , 
(F\Psi,F\Psi,F\Psi)^\eta_{G_\eta}\big\rangle
+\frac{\kappa^3}{24} \big\langle  F \Psi  , (\delta {G_\eta}, F\Psi,F\Psi,F\Psi)^\eta_{G_\eta}\big\rangle\no
=&\ \frac{\kappa^3}{6} \big\langle  B_\delta, D_\eta 
[ F\Psi, F\Xi (F\Psi,F\Psi,F\Psi)^\eta_{G_\eta} ]^\eta_{{G_\eta}}\big\rangle
-\frac{\kappa^3}{24} \big\langle B_\delta , D_\eta (F\Psi, F\Psi,F\Psi,F\Psi)^\eta_{G_\eta}\big\rangle,
\end{align}
where we used (\ref{FXCOMMU}), the cyclicity of the shifted dual string product, and $\delta {G_\eta} = D_\eta B_\delta$.
Thus we obtain
\begin{align}
E^{(4)}=\
-\frac{\kappa^3}{24} D_\eta (F\Psi, F\Psi,F\Psi,F\Psi)^\eta_{G_\eta}
+ \frac{\kappa^3}{6} D_\eta [ F\Psi, F\Xi (F\Psi,F\Psi,F\Psi)^\eta_{G_\eta} ]^\eta_{{G_\eta}}\,.
\label{EOM4}
\end{align}
Let us consider the invariance under the parameter $\Omega$ first.
The action is invariant 
if we can
determine $B_{\delta_\Omega}^{(4)}$ and $\delta_\Omega\Psi^{(3)}$ so that they satisfy
\begin{align}
0=& 
-\big\langle\!\big\langle \delta_{\Omega} \Psi^{(3)}, Y(Q\Psi+X\eta F\Psi)\big\rangle\!\big\rangle 
+ \big\langle D_\eta \Omega , E^{(4)}\big\rangle
+ \big\langle B_{\delta_{ \Omega} }^{(4)}, Q {G_\eta} \big\rangle.
\end{align}
However, since the second term vanishes,
\begin{equation}
 \langle D_\eta\Omega, E^{(4)}\rangle\ =\ \langle \Omega, D_\eta E^{(4)}\rangle\ =\ 0\,,
\end{equation}
%
we can consistently take
%
\begin{align}
B_{\delta_\Omega}^{(4)}=0\,,\qquad
\delta_\Omega\Psi^{(3)}=0\,.
\end{align}
%
Finally, let us consider the gauge invariances under the transformation with $\Lambda$.
We show that one can determine $\delta_\Lambda\Psi^{(3)}$ and $B_{\delta_\Lambda}^{(4)}$
so that the condition (\ref{fermion exp gen}) at quartic order, 
\begin{align}
0\
=&\
-\big\langle\!\big\langle {\delta_\Lambda}\Psi^{(1)}, YE^{(3)}\big\rangle\!\big\rangle 
+ \big\langle B_{{\delta_\Lambda}}^{(0)}, E^{(4)}\big\rangle
+ \big\langle B_{{\delta_\Lambda}}^{(2)}, E^{(2)}\big\rangle
\nonumber\\
&
-\big\langle\!\big\langle {\delta_\Lambda} \Psi^{(3)}, YE^{(1)}\big\rangle\!\big\rangle
+ \big\langle B_{{\delta_\Lambda}}^{(4)}, E^{(0)}\big\rangle
\label{GIR4Lamd}
\end{align}
holds, where the first three terms are already determined.
%
What we have to show is that these terms vanishes up to terms containing 
$E^{(1)}=Q\Psi+X\eta F\Psi$ and $E^{(0)}=Q{G_\eta}$, which can be compensated by appropriately determining
$\delta_\Lambda\Psi^{(3)}$ and $B_{\delta_\lambda}^{(4)}$, respectively:
\begin{align}
 0\ \cong&\ 
 \big\langle B_{{\delta_\Lambda}}^{(0)}, E^{(4)}\big\rangle
+ \big\langle B_{{\delta_\Lambda}}^{(2)}, E^{(2)}\big\rangle
-\big\langle\!\big\langle {\delta_\Lambda}\Psi^{(1)}, YE^{(3)}\big\rangle\!\big\rangle\,,
\label{quartic}
\end{align}
where $A\cong B$ denotes that $A$ equals to $B$ except for terms containing $E^{(1)}$ and $E^{(0)}$.
%
%
%
It is useful to note that
\begin{align}
QF\Psi\ &\cong\ 0\,,\\
Q[ B_1,..., B_n]^\eta_{{G_\eta}}\ &\cong\ 
\sum_{k=1}^n (-1)^{1+B_1 +... + B_{k-1}}[B_1, ..., QB_k, ... ,B_n]^\eta_{{G_\eta}}\,,\\
\{Q, D_\eta\}\ &\cong\ 0\,,\\
%
Q (F\Psi ,..., F\Psi )^\eta_{G_\eta}\ &\cong\ [ F\Psi ,..., F\Psi ]^\eta_{{G_\eta}}\,.
\end{align}
Utilizing them, we have
\begin{align}
- \llangle\delta_\Lambda\Psi^{(1)},YE^{(3)}\rrangle\
=&\  \frac{\kappa^3}{6}\big\langle  F \Xi  D_\eta [F\Psi, \Lambda]^\eta_{{G_\eta}}, 
X \eta F\Xi D_\eta(F\Psi,F\Psi,F\Psi)^\eta_{G_\eta}\big\rangle\no
%
=&\ \frac{\kappa^3}{6}\big\langle   [ F\Psi, \Lambda]^\eta_{{G_\eta}},  D_\eta F \Xi Q  
F\Xi D_\eta(F\Psi,F\Psi,F\Psi)^\eta_{G_\eta}\big\rangle\no
%
\cong&\ \frac{\kappa^3}{6}\big\langle   [ F\Psi, \Lambda]^\eta_{{G_\eta}},  
D_\eta F \Xi [F\Psi,F\Psi,F\Psi]^\eta_{{G_\eta}}\big\rangle
\nonumber\\
&
-\frac{\kappa^3}{6}\big\langle   [ F\Psi, \Lambda]^\eta_{{G_\eta}},  
Q  D_\eta F\Xi (F\Psi,F\Psi,F\Psi)^\eta_{G_\eta}\big\rangle\no
%
=&\
\frac{\kappa^3}{6}\big\langle   \Lambda,[F\Psi,[F\Psi,F\Psi,F\Psi]^\eta_{{G_\eta}}]^\eta_{{G_\eta}}\big\rangle
-\frac{\kappa^3}{6}\big\langle   F\Xi[ F\Psi, \Lambda]^\eta_{{G_\eta}},  
D_\eta[F\Psi,F\Psi,F\Psi]^\eta_{{G_\eta}}\big\rangle\no
&\hspace{30pt}
- \frac{\kappa^3}{6}\big\langle \Lambda, [F\Psi, QD_\eta F\Xi 
(F\Psi, F\Psi, F\Psi)^\eta_{G_\eta}]^\eta_{G_\eta}\big\rangle\,,
\end{align}
where we used $D_\eta F\Xi Q D_\eta F\Xi = (1-F\Xi D_\eta) Q D_\eta F\Xi \cong Q D_\eta F\Xi  $, and $\Xi^2=0$.
Similarly, one can show that the remaining two terms become
\begin{align}
\langle B_{\delta_\Lambda}^{(0)},E^{(4)}\rangle\
\cong&\ \frac{\kappa^3}{6} \big\langle \Lambda, 
[F\Psi,  QD_\eta F \Xi (F\Psi, F\Psi, F\Psi)^\eta_{G_\eta} ]^\eta_{{G_\eta}} \big\rangle
\nonumber\\
&
+\frac{\kappa^3}{24}\big\langle \Lambda,D_\eta [F\Psi,F\Psi,F\Psi,F\Psi]^\eta_{{G_\eta}}\big\rangle\,,\\
\langle B_{\delta_\Lambda}^{(2)},E^{(2)}\rangle\
\cong&\ \frac{\kappa^3}{4}\big\langle \Lambda,  
[F\Psi, F\Psi,[ F\Psi, F\Psi]^\eta_{{G_\eta}} ]^\eta_{{G_\eta}} \big\rangle 
\nonumber\\
&
-\frac{\kappa^3}{2}\big\langle F\Xi[ F\Psi, \Lambda]^\eta_{{G_\eta}}, 
[F\Psi, [F\Psi,F\Psi]^\eta_{{G_\eta}}]^\eta_{{G_\eta}}  \big\rangle\,.
\end{align}
Then we find (\ref{quartic}) holds 
by the $L_\infty$-relations of ${G_\eta}$-shifted dual products:
\begin{align}
&\frac{\kappa^3}{24}\big\langle   \Lambda, \Big(
D_\eta [F\Psi,F\Psi,F\Psi,F\Psi]^\eta_{{G_\eta}}
+4[F\Psi,[F\Psi,F\Psi,F\Psi]^\eta_{{G_\eta}}]^\eta_{{G_\eta}}
+6 [F\Psi, F\Psi,[ F\Psi, F\Psi]^\eta_{{G_\eta}} ]^\eta_{{G_\eta}} \Big)
\big\rangle \no
&\qquad-\frac{\kappa^3}{6}\big\langle   F\Xi[ F\Psi, \Lambda]^\eta_{{G_\eta}},  
\Big(
D_\eta[F\Psi,F\Psi,F\Psi]^\eta_{{G_\eta}}
+3[F\Psi, [F\Psi,F\Psi]^\eta_{{G_\eta}}]^\eta_{{G_\eta}} \Big) \big\rangle\ 
=\ 0\,.
\end{align}
%
By picking up the terms with $E^{(1)}$ and $E^{(0)}$, the transformations
$\delta_\Lambda\Psi^{(3)}$ and $B_{\delta_\Lambda}^{(4)}$ can be explicitly determined as
\begin{align}
B_{\delta_\Lambda}^{(4)}\ =&\
 -\frac{\kappa^4}{24} [(F\Psi, F\Psi,F\Psi,F\Psi)^\eta_{G_\eta}, \Lambda]^\eta_{{G_\eta}}
 +\frac{\kappa^4}{6} [[F\Psi , F\Xi ( F\Psi,F\Psi,F\Psi)^\eta_{G_\eta}]^\eta_{{G_\eta}}, \Lambda]^\eta_{{G_\eta}}\no
&+\frac{\kappa^4}{24} (F\Psi, F\Psi,F\Psi,F\Psi, D_\eta \Lambda)^\eta_{G_\eta}
-\frac{\kappa^4}{6} (F\Psi,F\Psi,F\Psi, F\Xi[F\Psi, D_\eta\Lambda]^\eta_{G_\eta})^\eta_{G_\eta}
\nonumber\\
& -\frac{\kappa^4}{6} [ F\Psi, F\Xi (F\Psi,F\Psi,F\Psi, D_\eta\Lambda)^\eta_{G_\eta}]^\eta_{{G_\eta}}
-\frac{\kappa^4}{6} [ F\Xi (F\Psi,F\Psi,F\Psi)^\eta_{G_\eta}, F\Psi, D_\eta \Lambda]^\eta_{{G_\eta}} \no
& +\frac{\kappa^4}{2} [F \Psi, F\Xi(F\Psi,F\Psi, F\Xi[F\Psi, D_\eta\Lambda]^\eta_{G_\eta})^\eta_{G_\eta}]^\eta_{{G_\eta}}
\nonumber\\
& 
+\frac{\kappa^4}{6} [F\Psi, F\Xi [F\Xi (F\Psi,F\Psi,F\Psi)^\eta_{G_\eta}, 
D_\eta \Lambda]^\eta_{{G_\eta}}]^\eta_{{G_\eta}}
\nonumber\\
&
+\frac{\kappa^4}{6} [F\Xi (F\Psi,F\Psi,F\Psi)^\eta_{G_\eta}, F\Xi[F\Psi, D_\eta\Lambda]^\eta_{G_\eta}]^\eta_{{G_\eta}},\\
\delta_\Lambda\Psi^{(3)}\ =&\
 -\frac{\kappa^3}{6} X \eta F \Xi D_\eta (F\Psi,F\Psi,F\Psi, D_\eta \Lambda)^\eta_{G_\eta}
 +\frac{\kappa^3}{2} X \eta F \Xi D_\eta (F\Xi[F\Psi, D_\eta\Lambda]^\eta_{{G_\eta}} ,F\Psi,F\Psi)^\eta_{G_\eta}\no
& +\frac{\kappa^3}{6} X \eta F \Xi D_\eta [F\Xi (F\Psi,F\Psi,F\Psi)^\eta_{G_\eta}, 
D_\eta \Lambda]^\eta_{{G_\eta}}\,.
\end{align}




\section{Summary and discussion}
\label{sec:discuss}
\setcounter{equation}{0}


Using the expansion in the number of the Ramond string field,
we have constructed in this paper a gauge invariant action
of heterotic string field theory at the quadratic and quartic order:
%
%
\begin{align}
S &=
-\frac{1}{2}\langle\!\langle \Psi,  YQ \Psi\rangle\!\rangle +\int^1_0 dt  
\big\langle B_t(t), Q {G_\eta}(t)+\frac{\kappa}{2} [F(t)\Psi,F(t)\Psi]^\eta_{G_\eta(t)}\big\rangle\no
&\qquad \qquad +\frac{\kappa^2}{24}\langle F\Psi, (F\Psi, F\Psi, F\Psi)^\eta_{G_\eta}\rangle
+ O(\Psi^6).
\label{action up to psi4}
\end{align}
This is invariant under the gauge transformations
with the parameter $\Lambda$,
\begin{align}
	B_{\delta_\Lambda}\ =&\ Q\Lambda 
+ \frac{\kappa^2}{2}[F\Psi,F\Psi,\Lambda]^\eta_{G_\eta}
-\kappa^2[F\Psi,F\Xi[F\Psi,\Lambda]^\eta_{G_\eta}]^\eta_{G_\eta}
\nonumber\\
&
 -\frac{\kappa^4}{24} [(F\Psi, F\Psi,F\Psi,F\Psi)^\eta_{G_\eta}, \Lambda]^\eta_{{G_\eta}}
 +\frac{\kappa^4}{6} [[F\Psi , F\Xi ( F\Psi,F\Psi,F\Psi)^\eta_{G_\eta}]^\eta_{{G_\eta}}, \Lambda]^\eta_{{G_\eta}}\no
&+\frac{\kappa^4}{24} (F\Psi, F\Psi,F\Psi,F\Psi, D_\eta \Lambda)^\eta_{G_\eta}
-\frac{\kappa^4}{6} (F\Psi,F\Psi,F\Psi, F\Xi[F\Psi, D_\eta\Lambda]^\eta_{G_\eta})^\eta_{G_\eta}
\nonumber\\
& -\frac{\kappa^4}{6} [ F\Psi, F\Xi (F\Psi,F\Psi,F\Psi, D_\eta\Lambda)^\eta_{G_\eta}]^\eta_{{G_\eta}}
-\frac{\kappa^4}{6} [ F\Xi (F\Psi,F\Psi,F\Psi)^\eta_{G_\eta}, F\Psi, D_\eta \Lambda]^\eta_{{G_\eta}} \no
& +\frac{\kappa^4}{2} [F \Psi, F\Xi(F\Psi,F\Psi, F\Xi[F\Psi, D_\eta\Lambda]^\eta_{G_\eta})^\eta_{G_\eta}]^\eta_{{G_\eta}}
\nonumber\\
& 
+\frac{\kappa^4}{6} [F\Psi, F\Xi [F\Xi (F\Psi,F\Psi,F\Psi)^\eta_{G_\eta}, 
D_\eta \Lambda]^\eta_{{G_\eta}}]^\eta_{{G_\eta}}
\nonumber\\
&
+\frac{\kappa^4}{6} [F\Xi (F\Psi,F\Psi,F\Psi)^\eta_{G_\eta}, 
F\Xi[F\Psi, D_\eta\Lambda]^\eta_{G_\eta}]^\eta_{{G_\eta}} + O(\Psi^6),\\
\delta_{\Lambda}\Psi\ =&\
- \kappa X\eta F\Xi D_\eta[F\Psi,\Lambda]^\eta_{G_\eta}
\nonumber\\
&
 -\frac{\kappa^3}{6} X \eta F \Xi D_\eta (F\Psi,F\Psi,F\Psi, D_\eta \Lambda)^\eta_{G_\eta}
 +\frac{\kappa^3}{2} X \eta F \Xi D_\eta (F\Psi,F\Psi, F\Xi[F\Psi, D_\eta\Lambda]^\eta_{{G_\eta}})^\eta_{G_\eta}\no
& +\frac{\kappa^3}{6} X \eta F \Xi D_\eta [F\Xi (F\Psi,F\Psi,F\Psi)^\eta_{G_\eta}, 
D_\eta \Lambda]^\eta_{{G_\eta}} + O(\Psi^5),
\end{align}
with the parameter $\Omega$,
\begin{align}
B_{\delta_\Omega}\ =&\ D_\eta\Omega + O(\Psi^6),\\
\delta_{\Omega}\Psi\ =&\ O(\Psi^5),
\end{align}
and with the parameter $\lambda$,
\begin{align}
B_{\delta_\lambda}\ =&\
-\kappa[ F\Psi ,F\Xi \lambda]^\eta_{G_\eta}
-\frac{\kappa^3}{6} (F\Psi,F\Psi,F\Psi,F\lambda)^\eta_{G_\eta}
+ \frac{\kappa^3}{6} [F\Xi (F\Psi,F\Psi,F\Psi)^\eta_{G_\eta},F \lambda]^\eta_{{G_\eta}} 
\nonumber\\
&
+\frac{\kappa^3}{2} [F \Psi, F\Xi (F\Psi,F\Psi,F\lambda)^\eta_{G_\eta}]^\eta_{{G_\eta}}
 + O(\Psi^6),\\
\delta_{\lambda}\Psi\ =&\ Q\lambda	
+ X\eta F\lambda
+ \frac{\kappa^2}{2} X\eta F \Xi D_\eta (F\Psi,F\Psi,F\lambda)^\eta_{G_\eta} + O(\Psi^5),\end{align}
except for the higher order in the Ramond string field.
The equations of motion derived from this action are 
\begin{align}
E^{NS}&=
QG_\eta+ \frac{\kappa}{2} [F\Psi,F\Psi]^\eta_{G_\eta}
\nonumber\\
&
-\frac{\kappa^3}{24}D_\eta\Big( 
(F\Psi, F\Psi,F\Psi,F\Psi)^\eta_{G_\eta}
- 4 [ F\Psi, F\Xi (F\Psi,F\Psi,F\Psi)^\eta_{G_\eta} ]^\eta_{G_\eta}\Big)
+ O(\Psi^6),\label{1002}\\
E^{R}&=Q\Psi+X\eta F\Psi+\frac{\kappa^2}{6} X \eta F\Xi D_\eta(F\Psi,F\Psi,F\Psi) ^\eta_{G_\eta}
+ O(\Psi^5).\label{1003}
\end{align}
Note that all of these results include all order terms
in the coupling constant $\kappa$ at each order
in the Ramond string field. %
We can also confirm that the action (\ref{action up to psi4}) reproduces
the four-point amplitudes with external fermions as given in the Appendix \ref{App B}.

The most important remaining task is to give a complete action and gauge transformation.
We finally discuss two observations which may provide clues to achieve it.
The first observation is a relation between the equations of motion and the gauge
transformations. At the beginning, it is natural to assume that the NS string field $V$ 
appears in the higher order action only in the form of $G_\eta$, since the corresponding 
ansatz is true for the case of the equations of motion in the dual formulation 
\cite{Kunitomo:2013mqa,Kunitomo:2014hba}. If we assume this ansatz the gauge transformation
with the parameter $\Omega$ does not subject to change any more.  
One can find that the gauge transformations $\delta_\Lambda\Psi$, $D_\eta B_{\delta_\lambda}$
and $\delta_\lambda\Psi$
are obtained by replacing fields in the equations of motion with gauge
parameters:\footnote{
One can see that similar relations hold exactly for the open superstring field theory.}
\begin{align}
\Big((D_\eta\Lambda) \frac{\delta}{\delta G_\eta}\Big) E^{(2k)}\ =&\ 
D_\eta B_{\delta_\Lambda}^{(2k)} + \kappa[E^{(2k)},\Lambda]^\eta_{G_\eta}\,,\label{951}
\qquad\textrm{for}\quad k=0,1,2\,,\\
\Big((D_\eta\Lambda) \frac{\delta}{\delta G_\eta}\Big) E^{(2k+1)}\ &\ =\delta_\Lambda\Psi^{(2k+1)},\label{954}
\qquad\textrm{for}\quad k=0,1\,,\\
-\Big(\lambda \frac{\delta}{\delta \Psi}\Big) E^{(2k)}\
=&\ D_\eta B_{\delta_\lambda}^{(2k)},\label{952}
\qquad \textrm{for}\quad k=1,2\,,\\
-\Big(\lambda \frac{\delta}{\delta \Psi}\Big) E^{(2k+1)}\ =&\ \delta_\lambda\Psi^{(2k+1)},\label{953}
\qquad\textrm{for}\quad k=0,1\,.
\end{align}
These relations might be an appearance of an $L_\infty$-structure, 
or equivalently a Batalin-Vilkovisky structure of the action:
in formulations based on the $L_\infty$-products,
the gauge transformation is given by a functional differentiation of the equation of motion.
To elucidate the role of theses relations in detail remains as future work which may
provide a hint to complete an action to all orders.

The second observation is the expression of the equations of motion obtained as a dual
form of the first-order equations of motion obtained in \cite{Kunitomo:2014hba}:
\begin{align}
(\eta+Q) \widetilde{B} +\sum_{m=2}^\infty\frac{1}{m!}[\widetilde{B} ^m]^\eta =0,\label{1012}
\end{align}
where $\widetilde{B} = \sum_{n=0}^\infty \widetilde B_{(n-2)/2}$. 
Expanding this in the picture number, the first two equations with the picture number
$P=-2$ and $-3/2$,
\begin{align}
 \eta\widetilde{B}_{-1} +
 \sum_{m=2}^\infty\frac{\kappa^{m-1}}{m!}[\widetilde{B}_{-1}^m]^\eta\ =&\ 0\,,\\
 D_\eta\widetilde{B}_{-1/2}\ =&\ 0\,,
\end{align} 
can be solved as
\begin{equation}
 \widetilde{B}_{-1}\ =\ G_\eta\,,\qquad \widetilde{B}_{-1/2}\ =\ F\Psi\,.
\end{equation}
The next two with $p=-1$ and $-1/2$,
\begin{align}
 QG_\eta + \frac{\kappa}{2}[\widetilde{B}_{-1/2},\widetilde{B}_{-1/2}]^\eta_{G_\eta}
 + D_\eta\widetilde{B}_0\ =&\ 0\,,\label{EOM1}\\
 Q\widetilde{B}_{-1/2} + \kappa [\widetilde{B}_0,\widetilde{B}_{-1/2}]^\eta_{G_\eta}
+ \frac{\kappa^2}{3!}[\widetilde{B}_{-1/2},\widetilde{B}_{-1/2},\widetilde{B}_{-1/2}]^\eta_{G_\eta}
+ D_\eta\widetilde{B}_{1/2}\ =&\ 0\,,\label{EOM2}
\end{align}
can be interpreted as equations of motion with the infinitely many subsidiary equations
determining the infinitely many ``auxiliary fields'', $\widetilde{B}_{n/2}\ (n\ge0)$. 
In the original formulation in \cite{Kunitomo:2014hba}, we can iteratively solve
these subsidiary conditions in the fermion expansion, and obtain the equations of motion.
We similarly assume here that the terms in the auxiliary fields with the lowest order in
fermion are $\widetilde{B}_{n/2}=O(\Psi^{n+4})$. Then the subsidiary equations simply
become
\begin{equation}
 Q\widetilde{B}_{n/2} + \frac{\kappa^{n+3}}{(n+4)!}
[\,\underbrace{F\Psi,F\Psi,\cdots,F\Psi}_{n+4}\,]^\eta_{G_\eta}\
=\ 0\,,
\end{equation}
which can be solved as
\begin{equation}
 \widetilde{B}_{n/2}^{(n+4)}\ =\ -\frac{\kappa^{n+3}}{(n+4)!}
(\,\underbrace{F\Psi, F\Psi, \cdots, F\Psi}_{n+4}\,)^\eta_{G_\eta}\,,
\label{sub1}
\end{equation}
except for the terms proportional to the lowest order equations of motion 
\begin{align}
 QG_\eta + \frac{\kappa}{2}[F\Psi, F\Psi]^\eta_{G_\eta}\ =\ 0\,,\\
 QF\Psi\ =\ 0\,,
\end{align}
obtained from (\ref{EOM1}) and (\ref{EOM2}), respectively.
Unfortunately, however, the next order equations of motion obtained by substituting
(\ref{sub1}) into (\ref{EOM1}) and (\ref{EOM2}) 
are 
not equivalent to our equations
of motion (\ref{1003}).  Although this difference can be filled by assuming that
$\widetilde{B}_{-1/2}$ contains the terms with the higher order in fermion as
\begin{equation}
 \widetilde{B}_{-1/2}\ =\ F\Psi -\frac{\kappa^2}{3!}D_\eta
  F\Xi(F\Psi,F\Psi,F\Psi)^\eta_{G_\eta} + O(\Psi^5),
\end{equation}
we cannot determine it without further assumption. 
Although this can be absorbed in the redefinition of the Ramond string field $\Psi$,
we have to find a way to reproduce the higher order terms in the equations of motion 
derived from the action, which may provides a clue for constructing a complete gauge 
invariant action.


\bigskip
\noindent
{\bf \large Acknowledgments}

\medskip
The authors would like to thank Yuji Okawa and Hiroaki Matsunaga for helpful discussion.

\appendix

\section{Construction of the dual gauge product}
\label{App A}
\setcounter{equation}{0}

In order to make this paper self-contained, we give in this appendix 
a construction of the dual string products. We follow the convention 
and notation of \cite{Goto:2015pqv}.

We first introduce the coalgebraic expression of string products, 
which is convenient to focus on their algebraic properties \cite{Erler:2014eba}.
The product of $n$ closed strings is described by a multilinear map $d_n:\mathcal H^{\wedge n}\to \mathcal H$,
where $\wedge$ is the {\it symmetrized tensor product} satisfying
$\Phi_1\wedge\Phi_2=(-1)^{\Phi_1\Phi_2}\Phi_2\wedge\Phi_1$.
This naturally induces a map from the {\it symmetrized tensor algebra} 
$\mathcal {S(H)}=\mathcal H^{\wedge 0}\oplus\mathcal H^{\wedge 1}\oplus\mathcal H^{\wedge 2}\oplus\cdots$
to $\mathcal S(\mathcal H)$ itself called a {\it coderivation}.
A coderivation ${\bf d}_n:\mathcal {S(H)}\to\mathcal {S(H)}$ is naturally derived from a map 
$d_n:{\cal H}^{\wedge n}\to\cal H$ as
\begin{align}
{\bf d}_n (\Phi _{1}\wedge \dots \wedge \Phi _{N})
&=(d_n\wedge\mathbb I_{N-n})(\Phi _{1}\wedge \dots \wedge \Phi _{N})\no
&= \sum_{\sigma}\frac{(-1)^{\sigma}}{n!(N-n)!} 
d_n(\Phi_{\sigma(1)},\cdots,\Phi_{\sigma(n)})\wedge \Phi_{\sigma(n+1)} \wedge \cdots \wedge\Phi_{\sigma(N)}\,,
\end{align}
for $\Phi _{1}\wedge \dots \wedge \Phi _{N} \in \mathcal H^{\wedge N\ge n}\subset  \mathcal{S(H)}$,
and it vanishes when acting on $ {\cal H}^{\wedge N< n}$.
The graded commutator of two coderivations ${\bf b}_{n}$ and ${\bf c}_{m}$, $[\![{\bf b}_n ,{\bf c}_m]\!]$,
is a coderivation derived from the map $[\![b_n,c_m]\!]:{\cal H}^{\wedge n+m-1}\to \cal H$
which is defined by 
\begin{align}
[\![b_n,c_m]\!]&=b_n(c_m\wedge\mathbb I_{n-1})-(-1)^{{\rm deg}(b_n){\rm deg}(c_m)}c_m(b_n\wedge\mathbb I_{m-1})\,.
\end{align}
Then the $L_\infty$-relation can be written as
\begin{align}
[\![ \mathbf L, \mathbf L]\!]=0\,,
\end{align}
where $\mathbf L = \mathbf L_1 + \mathbf L_2 + \mathbf L_3 + \cdots$ and $\mathbf L_k$ 
is a coderivation derived from the $k$-string product.

We introduce another map on the symmetrized tensor algebra, which is called a {\it cohomomorphism}.
From a set of multilinear maps $ \{ \mathsf f_n : \mathcal H^{\wedge n}\to \mathcal H'\}_{n=0}^{\infty}$,
one can naturally define a cohomomorphism $\widehat{\mathsf f} : \mathcal{S(H)}\to \mathcal{S(H')}$,
which acts on $\Phi _{1}\wedge \dots \wedge \Phi _{n} \in \mathcal H^{\wedge n}\subset  \mathcal{S(H)}$ as
\begin{align}
\widehat{\mathsf f} (\Phi _{1}\wedge \dots \wedge \Phi _{n})\ =\
\sum_{i \leq n} \sum_{k_{1} < \dots < k_{i}=n} 
&e^{\wedge \mathsf f_0}\wedge{\sf f}_{k_{1}} ( \Phi _{1} , \dots , \Phi _{k_{1}}) 
\wedge {\sf f}_{k_{2}-k_{1}} ( \Phi _{k_{1} +1} , \dots , \Phi _{k_{2}} ) \wedge \no
&\hspace{15pt}\dots \wedge  {\sf f}_{k_{i} -k_{i-1}} ( \Phi _{k_{i-1}+1} , \dots , \Phi _{n} )\,.
\end{align}
We also introduce a projector $\pi_1$ from the symmetrized tensor algebra to the single-state
space, 
$\mathcal {S(H)}\to\mathcal {H}$, as
\begin{align}
\pi _{1} \big( \Phi _{0} + \Phi _{1} \wedge \Phi _{2} + \Phi _{3} \wedge \Phi _{4} \wedge
 \Phi _{5} + \dots \big)\
 =\ \Phi _{0}\,. 
\end{align}

The NS string products for heterotic string field theory, 
$\mathbf L^\mathrm{NS}(\tau)= \sum_{p=0}^{\infty} \tau^p \mathbf L_{p+1}$,
had been constructed in \cite{Erler:2014eba}.
They satisfy the $L_\infty$-relation 
$[\![\mathbf L^\mathrm{NS}{(\tau)},\mathbf L^\mathrm{NS}{(\tau)}]\!]=0$,
cyclicity, and (graded) commutativity with $\eta$: $[\![\eta,\mathbf L^\mathrm{NS}(\tau)]\!]=0$.
The $(p+1)$-product $\mathbf L^\mathrm{NS}_{p+1}$ carries the ghost number $1-2p$ and the picture number $p$.
The whole string product $\mathbf L^\mathrm{NS}(\tau)$ is given 
by a similarity transformation of the BRST operator $\mathbf Q$ as
\begin{align}
{\bf L}^\mathrm{NS}(\tau) &= \widehat{\bf G}^{-1}(\tau){\bf Q} \widehat{\bf G}(\tau)\,,
\label{L NS}
\end{align}
where
$\widehat{\mathbf G}(\tau)$ is an invertible cohomomorphism given by the path-ordered exponential map:
\begin{align}
\widehat{\bf G}(\tau)=\overset{\leftarrow}{\mathcal P} 
\exp \left( \int_{0}^{\tau} d\tau' {\boldsymbol \uplambda^{[0]}}(\tau')\right)\,,
\label{G}\qquad
\widehat{\bf G}^{-1}(\tau)
=\overset{\rightarrow}{\mathcal P} \exp \left(-\int_{0}^{\tau} d\tau' {\boldsymbol \uplambda^{[0]}}(\tau')\right)\,,
\end{align}
where $\boldsymbol \uplambda^{[0]}(\tau)=\sum_{p=0}^\infty \tau^p \boldsymbol
\uplambda^{[0]}_{p+2}$, called gauge products, can be determined iteratively.
The arrow $\leftarrow$\ ($\rightarrow$) on ${\mathcal P}$ denotes that the operator at later time acts from the right (left).
The $(p+2)$-gauge product $\boldsymbol \uplambda^{[0]}_{p+2}$ carries ghost number $-2(p+1)$ and picture number $p+1$.
%
The cohomomorphisms
$\widehat{\mathbf G}(\tau)$ and $\widehat{\mathbf G}^{-1}(\tau)$ satisfy
\begin{align}
\partial_{\tau}\widehat{\mathbf G}(\tau)=\widehat{\mathbf G}(\tau)  \boldsymbol \uplambda^{[0]}(\tau)\,,\qquad
\partial_{\tau}\widehat{\mathbf G}^{-1}(\tau)=-\boldsymbol \uplambda^{[0]}(\tau)\widehat{\mathbf G}^{-1}(\tau)\,.
\label{deq g}
\end{align}
The $L_\infty$-relations are followed from the nilpotency of $\mathbf{Q}$ as
\begin{align}
[\![ \mathbf{L}^{\mathrm{NS}}(\tau), \mathbf{L}^{\mathrm{NS}}(\tau)]\!]\ =&\
2\big({\bf L}^\mathrm{NS}(\tau)\big)^2 \nonumber\\
=&\ 2 \widehat{\bf G}^{-1} (\tau){\bf Q}^2
\widehat{\bf G}(\tau)
=\ 0\,.
\end{align}
%
%
The cyclicity and commutativity 
$[\![\boldsymbol \upeta,\mathbf L^\mathrm{NS}(\tau)]\!]=0,$
is realized by a suitable choice of an initial gauge product $\uplambda^{[0]}$,
whose explicit example is given in \cite{Erler:2014eba}.

Under these preparations,
we summarize the construction of the dual string products given in \cite{Goto:2015pqv}.
We introduce a coderivation $\mathbf L^\eta(\tau)= \sum_{p=0}^{\infty} \tau^p \mathbf L^\eta_{p+1}$
which provides a set of the dual string products by
\begin{align}
[V_1,V_2,...,V_n]^\eta = \pi_1 \mathbf L^\eta_n (V_1\wedge V_2 \wedge ... \wedge V_n)\,.
\end{align}
This $n$-th dual product $\mathbf L^\eta_n$ carries ghost number $3-2n$ and picture number
$n-2$ as expected.
The whole coderivation $\mathbf L^\eta(\tau)$ is degree odd, and 
can be constructed using the cohomomorphism ${\widehat{\mathbf G}}(\tau)$
appearing in 
the NS product 
$\mathbf L^\mathrm{NS}(\tau)=\widehat{{\mathbf G}}^{-1}(\tau) \mathbf Q \widehat{{\mathbf G}}(\tau)$ by
\begin{align}
\mathbf L^\eta(\tau) = \widehat{\mathbf G}(\tau) \boldsymbol\upeta \widehat{\mathbf G}^{-1}(\tau)\,.
\end{align}
By construction, they satisfy the $L_\infty$-relation
\begin{align}
[\![\mathbf L^\eta(\tau),\mathbf L^\eta(\tau)]\!]
=0\,.
\end{align}
The anti-commutativity $[\![ \mathbf Q, \mathbf L^\eta(\tau)]\!]=0$ follows from 
$[\![\boldsymbol \upeta, \mathbf L^{\rm NS}(\tau)]\!]=0$ as
\begin{align}
[\![\mathbf Q, \mathbf L^\eta(\tau)]\!]
=[\![\mathbf Q, \widehat{\mathbf G} (\tau) \boldsymbol\upeta \widehat{\mathbf G}^{-1}(\tau)]\!]
=\widehat{\mathbf G} (\tau)[\![ \mathbf L^{\rm NS}(\tau),\boldsymbol\upeta ]\!]\widehat{\mathbf G}^{-1}(\tau)
=0\,.
\end{align}
The cyclicity of $L^\eta(\tau)$ again follows from that of the gauge product $\boldsymbol \uplambda^{[0]}$.
We can give an explicit expression of the dual string product using the bosonic string
product if necessary, for example, 
\begin{align}
[V_1,V_2]^\eta &=-[V_1,V_2],\\
%
[V_1,V_2,V_3 ]^\eta\ =&\ -\frac{1}{4}\Bigg(
X_0[V_1,V_2,V_3] 
+[X_0V_1,V_2,V_3]+[V_1,X_0V_2,V_3]+[V_1,V_2,X_0V_3]
\nonumber\\
&\hspace{35mm}
+(-1)^{V_1}\xi_0 [V_1,[V_2,V_3]]
-(-1)^{V_1}[\xi_0 V_1,[V_2,V_3]]
\nonumber\\
&\hspace{50mm}
+[V_1,[\xi_0 V_2,V_3]]
+(-1)^{V_2}[V_1,[V_2, \xi_0 V_3]]
\nonumber\\
&\hspace{10mm}
+(-1)^{V_1(V_2+V_3)}\Big(
(-1)^{V_2}\xi_0 [V_2,[V_3,V_1]]
-(-1)^{V_2}[\xi_0 V_2,[V_3,V_1]]
\nonumber\\
&\hspace{50mm}
+[V_2,[\xi_0 V_3,V_1]]
+(-1)^{V_3}[V_2,[V_3, \xi_0 V_1]]\Big)
\nonumber\\
&\hspace{10mm}
+(-1)^{V_3(V_1+V_2))}\Big(
(-1)^{V_3}\xi_0 [V_3,[V_1,V_2]]
-(-1)^{V_3}[\xi_0 V_3,[V_1,V_2]]
\nonumber\\
&\hspace{50mm}
+[V_3,[\xi_0 V_1,V_2]]
+(-1)^{V_1}[V_3,[V_1, \xi_0 V_2]]\Big)\Bigg)\,,
\end{align}
where $X_0=\{Q,\xi_0\}$.
These dual string products are defined on the basis of $\mathbf{L}^{NS}(\tau)$ in the NS
sector. However, we can extend it to include
the Ramond sector simply by considering $V_i$ is either the NS string field
or the Ramond string field, which preserves the necessary properties,
the $L_\infty$-relation, cyclicity and the commutativity with $\mathbf{Q}$,
to construct the gauge invariant action. 
It should be emphasized here that it is not necessary to introduce any special
picture changing operator only for the Ramond sector to define the dual string products.

The dual string product $\mathbf L^\eta$ is commutative with $\mathbf{Q}$, and
its second derivative with respect to $\tau$
can be written as the commutator of $\mathbf{Q}$ and a product $\boldsymbol\uprho$:
\begin{align}
\partial_\tau^2\mathbf L^\eta(\tau)\ 
=\ [\![ \mathbf Q, \boldsymbol\uprho (\tau)]\!]\
=\ \sum_{n=0}^\infty \tau^n [\![ \mathbf Q, \boldsymbol\uprho_{n+3}]\!]\,.
\end{align}
The dual gauge products can be read from $\boldsymbol\uprho$ as
\begin{align}
(V_1, V_2, ..., V_n)^\eta\ =\ \frac{1}{(n+1)(n+2)}\pi_1 
\boldsymbol \uprho_n (V_1 \wedge V_2 \wedge ... \wedge V_n)\,,\qquad (n\ge3)\,.
\end{align}
In order to obtain an explicit expression of $\boldsymbol{\uprho}$,
we introduce a coderivation $\mathbf L^{[1]} (\tau) = \sum_{n=0}^\infty \tau^n\mathbf L^{[1]}_{n+2}$
which is an intermediate products with deficit picture $1$ given in \cite{Erler:2014eba}.
It is related
to the gauge products $\boldsymbol\uplambda^{[0]}(\tau)$ as
\begin{align}
[\![ \boldsymbol\upeta ,\boldsymbol\uplambda^{[0]}(\tau)]\!]\ =\ \mathbf L^{[1]}(\tau)\,,
\end{align}
and satisfies $\mathbf L^{[1]}(\tau=0)= \mathbf L_2^\mathrm{B}$,
where $\mathbf L_2^\mathrm{B}$ is a coderivation derived from the simple two-string product 
for closed string 
without any insertion of superconformal ghost. 
%
We also introduce a coderivation
$\boldsymbol \uplambda^{[1]}(\tau)=\sum_{n=0}^\infty \tau^n\boldsymbol\uplambda^{[1]}_{n+3}$
derived from a set of intermediate gauge products with deficit picture $1$ \cite{Erler:2014eba}.
It satisfies the relation,
\begin{align}
\partial_\tau \mathbf L^{[1]}(\tau) =
[\![\mathbf L^{[1]}(\tau), \boldsymbol \uplambda^{[0]}(\tau)]\!]
+[\![\mathbf L^{[0]}(\tau), \boldsymbol \uplambda^{[1]}(\tau)]\!]\,,\label{442}
\end{align}
with $\mathbf{L}^{[0]}=\mathbf{L}^{NS}$.
Then, utilizing these products and 
their path-ordered exponential maps,
we can rewrite $\mathbf L^\eta$ as
\begin{align}
\mathbf L^\eta (\tau) 
&= \widehat{\mathbf G}(\tau) \boldsymbol\upeta \widehat{\mathbf G}^{-1} (\tau)\no
&= \boldsymbol\upeta + \widehat{\mathbf G}(\tau) [\![ \boldsymbol\upeta ,\widehat{\mathbf G}^{-1}(\tau)]\!]\no
&= \boldsymbol\upeta - \int^\tau_0 d\tau'\widehat{\mathbf G} (\tau') 
\mathbf L^{[1]}(\tau') \widehat{\mathbf G}^{-1}(\tau')\,.\label{LETA LDEF1}
\end{align}
The integrand in the second term becomes
\begin{align}
\widehat{\mathbf G} (\tau') \mathbf L^{[1]}(\tau') \widehat{\mathbf G}^{-1}(\tau')
&=\widehat{\mathbf G} (0) \mathbf L^{[1]}(0) \widehat{\mathbf G}^{-1}(0)
+\int^{\tau'}_0 d\tau'' \; \partial_{\tau''}\Big( \widehat{\mathbf G} (\tau'') 
\mathbf L^{[1]}(\tau'') \widehat{\mathbf G}^{-1}(\tau'')\Big) \no
&=\mathbf L^\mathrm{B}_2
+\int^{\tau'}_0 d\tau'' 
[\![ \mathbf Q , \widehat{\mathbf G} (\tau'')
\boldsymbol \uplambda^{[1]}(\tau'')\widehat{\mathbf G}^{-1}(\tau'')]\!]\,,
\end{align}
where we used $\widehat{\mathbf{G}}(0)=\mathbb{I}$,
$\mathbf{L}^{[1]}(0)=\mathbf{L}_2^{\textrm{B}}$, and
\begin{align}
\partial_{\tau''}\Big( \widehat{\mathbf G} (\tau'') \mathbf L^{[1]}(\tau'') \widehat{\mathbf G}^{-1}(\tau'')\Big)\ 
=&\ \widehat{\mathbf G} (\tau'') [\![ \mathbf L^{[0]}(\tau''),
 \boldsymbol\uplambda^{[1]}(\tau'')]\!] \widehat{\mathbf G}^{-1}(\tau'')
\nonumber\\
=&\ [\![ \mathbf Q , \widehat{\mathbf G} (\tau'')\boldsymbol \uplambda^{[1]}(\tau'')\widehat{\mathbf G}^{-1}(\tau'')]\!]\,,
\end{align} 
which follows from 
the differential equations (\ref{deq g}) and (\ref{442}), and (\ref{L NS}) with $\mathbf{L}^{[0]}=\mathbf{L}^{NS}$.
Then eventually the dual string products $\mathbf L^\eta$ can be written as
\begin{align}
\mathbf L^\eta (\tau) 
&= \boldsymbol\upeta - \int^\tau_0 d\tau'\Big(\mathbf L^\mathrm{B}_2
+\int^{\tau'}_0 d\tau'' [\![ \mathbf Q , \widehat{\mathbf G} (\tau'')
\boldsymbol \uplambda^{[1]}(\tau'')\widehat{\mathbf G}^{-1}(\tau'')]\!]\Big)\no
&=\boldsymbol \upeta - \tau \mathbf L^{\rm B}_2 - \int^{\tau}_0 d\tau''(\tau-\tau'')  
[\![\mathbf Q, \widehat{\mathbf G} (\tau'')\boldsymbol\uplambda^{[1]}(\tau'')\widehat{\mathbf G}^{-1}(\tau'')]\!]\,.
\label{L eta}
\end{align}
%
%
%
%
%
Here we used $\int^\tau_0 d\tau' \int^{\tau'}_0 d\tau'' = \int_0^\tau d\tau'' 
\int^\tau_{\tau''}d\tau'$ and carried out $\tau'$-integral.
In this expression, the commutativity $[\![\mathbf Q, \mathbf L^\eta ]\!]=0$ is manifest.
Differentiating (\ref{L eta}) with respect to $\tau$, we obtain
\begin{align}
\partial_\tau  \mathbf L^\eta (\tau)
&=- \mathbf L^{\rm B}_2 - \int^{\tau}_0 d\tau''  [\![\mathbf Q, 
\widehat{\mathbf G} (\tau'')\boldsymbol\uplambda^{[1]}(\tau'')\widehat{\mathbf G}^{-1}(\tau'')]\!]\,,\\
\partial_\tau^2  \mathbf L^\eta (\tau)
&= -\ [\![\mathbf Q, \widehat{\mathbf G} (\tau)\boldsymbol\uplambda^{[1]}(\tau)
\widehat{\mathbf G}^{-1}(\tau)]\!]\,.\label{DE LETA Q EXACT}
\end{align}
Therefore we can define the product $\boldsymbol{\uprho}$ by two gauge products
$\boldsymbol{\uplambda}^{[0]}$
and $\boldsymbol{\uplambda}^{[1]}$ as
\begin{align}
\boldsymbol\uprho(\tau) = - \mathbf G(\tau) 
\boldsymbol\uplambda^{[1]}(\tau)\mathbf G^{-1}(\tau)\,.\label{C17C}
\end{align}
The cyclicity of $\boldsymbol\uprho$ 
follows from that of 
$\boldsymbol\uplambda^{[0]}$ and $\boldsymbol\uplambda^{[1]}$.
Expanding  (\ref{C17C}) in powers of $\tau$,
we obtain the following expressions for the first few orders:
\begin{align}
\boldsymbol\uprho_3 &= -\ \boldsymbol\uplambda^{[1]}_3\,,\\
\boldsymbol\uprho_4 &= -\ \Big(\boldsymbol\uplambda^{[1]}_4 
+ [\![ \boldsymbol\uplambda^{[0]}_2 ,\boldsymbol\uplambda^{[1]}_3]\!]\Big)\,,\\
\boldsymbol\uprho_5 &= -\ \Big(
\boldsymbol\uplambda^{[1]}_5 
+ [\![ \boldsymbol\uplambda^{[0]}_2 ,\boldsymbol\uplambda^{[1]}_4]\!]
+\frac{1}2 [\![\boldsymbol\uplambda^{[0]}_3,\boldsymbol\uplambda^{[1]}_3]\!] 
+\frac12 [\![\boldsymbol\uplambda^{[0]}_2,[\![ \boldsymbol\uplambda^{[0]}_2 ,\boldsymbol\uplambda^{[1]}_3]\!]]\!]
\Big)\,.
\end{align}
The explicit form of three-string dual gauge product is, for example, given by 
\begin{align}
(V_1, V_2, V_3)^\eta\ =&\ -\frac14\Big( \xi[V_1,V_2,V_3]-[\xi V_1,V_2,V_3]
\nonumber\\
&\hspace{32mm}
-(-1)^{V_1}[V_1,\xi V_2,V_3]-(-1)^{V_1+V_2}[V_1,V_2,\xi V_3]\Big)\,.
\label{dual 3gp}
\end{align}

\vspace{10mm}

\section{Four-point amplitudes with external fermions}
\label{App B}
\setcounter{equation}{0}

In this appendix, we illustrate how the on-shell physical amplitudes
with external fermions are reproduced 
from the heterotic string field theory
by concentrating on the case of four-point amplitudes which can be calculated 
only from the action up to $\mathcal{O}(\Psi^4)$ constructed in this paper.
We follow the notations and conventions in \cite{Kunitomo:2015hda}. 

\subsection{Propagators and vertices}

The kinetic terms of the NS string field $V$ and
the Ramond string field $\Psi
$ are obtained from (\ref{dual action NS}) and (\ref{quadratic action}) as
\begin{equation}
 S_0\ 
=\ \frac{1}{2}\langle\eta V, QV\rangle
-\frac{1}{2}\langle \xi_0\Psi, YQ\Psi\rangle\,.
\end{equation}
This is invariant under the gauge transformations
\begin{equation}
 \delta V\ =\ Q\Lambda + \eta\Omega\,,\qquad
 \delta \Psi\ =\ Q\lambda\,,
\end{equation}
which we fix here by gauge conditions
\begin{align}
 b_0^+V\ =\ \xi_0 V\ =\ 0\,,\qquad
 b_0^+\Psi\ =\ 0\,.
\label{gauge conditions}
\end{align}
Under these gauge conditions the propagators
of the NS and the Ramond string fields can be found as
\begin{align}
 \overbracket[0.5pt]{\!\!\!\!|V\rangle\langle V|\!\!\!\!}\ =&\ 
\xi_0\frac{b_0^-b_0^+}{L_0^+}\delta(L_0^-)
\nonumber\\
=&\ \int d^2 T\,(\xi_0b_0^-b_0^+)\,
e^{-TL_0^+-i\theta L_0^-}\ \equiv\ \Pi_{NS}\,,
\label{NS propagator}\\
 \overbracket[0.5pt]{\!\!\!\!|\Psi\rangle\langle\Psi|\!\!\!\!}\ =&\ 
-\frac{b_0^-b_0^+ X\eta}{L_0^+}\delta(L_0^-)
\nonumber\\
=&\ - \int d^2 T\,(b_0^-b_0^+X\eta)\, e^{-TL_0^+-i\theta L_0^-}\
\equiv\ \Pi_{R}\,,
\label{R propagator}
\end{align}
respectively, where $\int d^2 T=\int_0^\infty dT\int_0^{2\pi} \frac{d\theta}{2\pi }$\,.
Notice that the Ramond propagator satisfies\footnote{
The condition for the BPZ conjugate, the latter equation in (\ref{const prop}), 
can  be understood
from the inner product between restricted states $|\Psi_1\rangle$ and $|\Psi_2\rangle$
in the large Hilbert space: $\langle\Psi_1|\xi_0 Y|\Psi_2\rangle=
\langle\Psi_1|\xi_0YXY|\Psi_2\rangle=\langle\Psi_1|(\xi_0YX\eta)\xi_0Y|\Psi_2\rangle$.
}
\begin{align}
 \eta\Pi_R\ =&\ \Pi_R\eta\ =\ 0\,,\label{const prop small}\\
 XY\Pi_R =&\ \Pi_R(\xi_0YX\eta)\ =\ \Pi_R\,,\label{const prop}
\end{align}
which expresses that 
only the Ramond states satisfying the constraint (\ref{constR1}) propagate.

The necessary interaction vertices can be read from (\ref{dual action NS}), (\ref{quadratic action})
and (\ref{quartic action}) by expanding them in the coupling constant $\kappa$:
\begin{align}
 S_1^{(0)}\ =&\ \frac{\kappa}{3!}\langle\eta V, [V, QV]\rangle\,,
\label{NS3}\\
S_2^{(0)}\ =&\ \frac{\kappa^2}{4!}\langle\eta V,[V,(QV)^2]\rangle
+ \frac{\kappa^2}{4!}\langle\eta V, [V, [V, QV]\rangle\,,
\label{NS4}\\
S_1^{(2)}\ =&\ -\frac{\kappa}{2}\langle\Psi, [V,\Psi]\rangle\,,
\label{Yukawa}\\
S_2^{(2)}\ =&\ 
-\frac{\kappa^2}{16}\Big(
\langle\xi_0\Psi, [QV, \eta V, \Psi]\rangle
+\langle\Psi,[\xi_0 QV, \eta V, \Psi]\rangle
\nonumber\\
&
+ \langle\Psi, [QV, \xi_0\eta V, \Psi]\rangle
+\langle\Psi, [QV, \eta V, \xi_0\Psi]\rangle\Big)
-\frac{\kappa^2}{2}\langle\Xi[\eta V, \Psi], [V, \Psi]\rangle\,,
\label{NS2R2}
\\
S_2^{(4)}\ =&\ \frac{\kappa^2}{4!}\langle\xi_0\Psi,[\Psi^3]\rangle\,.
\label{R4}
\end{align}
From these propagators and vertices, we can uniquely calculate
the tree-level four-point amplitudes
including 
external fermions.

\subsection{Four-fermion amplitude}

Let us first consider the amplitude with four external fermions in the Ramond sector.
The contributions to the four-fermion amplitude come from
the four Feynman diagrams, the $s$-, $t$-, $u$-channel diagrams
and a contact type diagram containing a four-string vertex.
Following the convention in \cite{Kunitomo:2015hda} we label each external string
with $A$, $B$, $C$, and $D$, and denote the $s$-, $t$-, and $u$-channel
contributions as $(AB|CD)$, $(AC|BD)$, and $(AD|BC)$, respectively.
Then the $s$-channel contribution can be calculated from the NS propagator
(\ref{NS propagator}) and the three-string interaction (\ref{Yukawa}) as
\begin{align}
 \mathcal{A}^{(AB|CD)}_{F^4}\ =&\
(-\kappa)^2 \langle\Psi_A\Psi_B\,\,\,
 \overbracket[0.5pt]{\!\!\!V\rangle\langle V\!\!\!}\,\,\,\Psi_C\Psi_D\rangle
\nonumber\\
=&\ \kappa^2\int d^2T_s
\langle\Psi_A\Psi_B(\xi_0b_0^- b_0^+)\Psi_C\Psi_D\rangle_{W_s}
\nonumber\\
=&\ \kappa^2\int d^2T_s\llangle \Psi_A\Psi_B(b_0^- b_0^+)\Psi_C\Psi_D\rrangle_{W_s}\,,
\end{align}
where we denoted the moduli of the $s$-channel propagator as $(T_s,\theta_s)$ and  
the correlation $\langle\cdots\rangle_{W_s}$,
or $\llangle\cdots\rrangle_{W_s}$, is evaluated as the conformal field theory on
the $s$-channel string diagram depicted in \cite{Kunitomo:2013mqa}.
The inserted $\xi_0$ and $b^\pm_0$ are the zero mode with respect to 
the local coordinate on the NS propagator.\footnote{
These are denoted $\xi_c$ and $b^\pm_c$ in \cite{Kunitomo:2013mqa}.}
The symbols $\Psi_A,\cdots,\Psi_D$ represent the wave functions of 
the Ramond external states, which satisfy the constraints 
(\ref{constR1}), gauge condition (\ref{gauge conditions}) and 
the on-shell condition $Q\Psi=0$\,. They can be given by the specific 
vertex operators if necessary.

In this notation the $t$- and $u$-channel contributions can similarly be calculated as
\begin{align}
 \mathcal{A}^{(AC|BD)}_{F^4}\ 
=&\ \kappa^2\int d^2T_t\llangle \Psi_A\Psi_C(b_0^- b_0^+)\Psi_B\Psi_D\rrangle_{W_t}\,,\\
 \mathcal{A}^{(AD|BC)}_{F^4}\ 
=&\ \kappa^2\int d^2T_u\llangle \Psi_A\Psi_D(b_0^- b_0^+)\Psi_B\Psi_C\rrangle_{W_u}\,.
\end{align}

A contact type interaction (\ref{R4}) gives the other contribution, denoted $(ABCD)$,
which fills the gap in the moduli integration \cite{Saadi:1989tb}:
\begin{equation}
 \mathcal{A}^{(ABCD)}_{F^4}\ =\ \kappa^2\int d\theta_1d\theta_2
\llangle (b_{C_1}b_{C_2}) \Psi_A\Psi_B\Psi_C\Psi_D\rrangle_{W_4}\,,
\label{four fermion contact}
\end{equation}
where the integration parameters $\theta_1$ and $\theta_2$ determine the shape
of a tetrahedron along which four strings in the vertex (\ref{R4}) are glued,
and $b_{C_{1,2}}$ denote the corresponding anti-ghost insertions, the details of which
are given in \cite{Kugo:1989tk}.

Summing up all these contributions
the on-shell four-fermion amplitude becomes
\begin{align}
 \mathcal{A}_{F^4}\ =&\ 
\mathcal{A}^{(AB|CD)}_{F^4}
+ \mathcal{A}^{(AC|BD)}_{F^4}
+ \mathcal{A}^{(AD|BC)}_{F^4}
+ \mathcal{A}^{(ABCD)}_{F^4}
\nonumber\\
=&\
\kappa^2 \int d^2 T_s\,\llangle \Psi_A\Psi_B(b_0^- b_0^+)\Psi_C\Psi_D\rrangle_{W_s}
+ \kappa^2 \int d^2 T_t\,\llangle \Psi_A\Psi_C(b_0^- b_0^+)\Psi_B\Psi_D\rrangle_{W_t}
\nonumber\\
&\
+ \kappa^2 \int d^2 T_u\,\llangle \Psi_A\Psi_D(b_0^- b_0^+)\Psi_B\Psi_C\rrangle_{W_u}
+ \kappa^2\int d^2 \theta\,
\llangle (b_{C_1}b_{C_2}) \Psi_A\Psi_B\Psi_C\Psi_D\rrangle_{W_4}\,.
\label{Four fermion}
\end{align}
From the fact that the bosonic closed string field theory
reproduces the correct perturbative amplitudes,
we can conclude that the amplitude (\ref{Four fermion}) agrees with that
obtained in the first quantized formulation.

\subsection{Two-fermion-two-boson amplitude}

We can similarly calculate the two-fermion-two-boson amplitude.
Suppose strings $A$ and $B$ are fermions in the Ramond sector and
strings $C$ and $D$ are bosons in the NS sector. The $s$-channel
contributions can be calculated using the NS propagator (\ref{NS propagator})
and the vertices (\ref{NS3}) and (\ref{Yukawa}):
\begin{align}
 \mathcal{A}^{(AB|CD)}_{F^2B^2}\ =&\ 
-\frac{\kappa^2}{2}\langle\Psi_A\Psi_B\,\,\, 
\overbracket[0.5pt]{\!\!\! V\rangle\langle V\!\!\!}\,\,\,
\big((QV_C)\, (\eta V_D) + (\eta V_C)\, (QV_D)\big)\rangle_{W_s}\nonumber\\
=&\ -\frac{\kappa^2}{2}\int d^2T_s\,
\langle\Psi_A\Psi_B (\xi_0 b_0^- b_0^+) 
\big((QV_C)\, (\eta V_D) + (\eta V_C)\, (QV_D)\big)\rangle_{W_s}
\nonumber\\
=&\ -\frac{\kappa^2}{2}\int d^2T_s\,
\llangle\Psi_A\Psi_B (b_0^- b_0^+) 
\big((X_0\eta V_C)\ (\eta V_D) + (\eta V_C)\ (X_0\eta V_D)\big)\rrangle_{W_s}\,,
\label{F2B2 s}
\end{align}
where $V_C$ and $V_D$ represent the wave functions of the NS external states,
which satisfy the gauge conditions (\ref{gauge conditions}) and the on-shell condition
$Q\eta V=0$\,. In the last equation, we used $QV=X_0\eta V$ for this wave function $V$\,.

The $t$-channel contribution in this case is calculated using the Ramond propagator
(\ref{R propagator}) and the vertex (\ref{Yukawa}):
\begin{align}
 \mathcal{A}^{(AC|BD)}_{F^2B^2}\ =&\ 
-\kappa^2
\langle\Psi_A V_C\,\,\, 
\overbracket[0.5pt]{\!\!\!\Psi\rangle\langle\Psi\!\!\!}\,\,\,
\Psi_B V_D\rangle
\nonumber\\
=&\ -\kappa^2\int d^2T_t\,
\langle\Psi_A V_C (b_0^- b_0^+ X\eta) \Psi_B V_D\rangle_{W_t}\,.
\end{align}
In order to smoothly connect the contributions of
the four diagrams at their boundaries and sum up them to the integration
over the whole moduli space,
we rearrange the integrand so as to be the correlation function 
of the same external vertices, 
$\Psi_A$, $\Psi_B$, $((X_0\eta V_C)\,(\eta V_D)+(\eta V_C)\,(X_0\eta V_D))/2$,
and the operator insertion $(b_0^-b_0^+)$ as those in (\ref{F2B2 s}).
In particular we move the picture changing operator $X$ in the Ramond propagator, 
whose form is highly depend on the coordinate system of the propagator,
to an external state by using the relation $X=\{Q,\Xi\}$.
After a little manipulation, we obtain
\begin{align}
\mathcal{A}^{(AC|BD)}_{F^2B^2}\ =&\ -\frac{\kappa^2}{2}\int d^2T_t\,
\Big(\langle\Psi_A V_C (b_0^- b_0^+ X) \Psi_B\, (\eta V_D)\rangle_{W_t}
+\langle\Psi_A\, (\eta V_C) (b_0^- b_0^+ X) \Psi_B V_D\rangle_{W_t}\Big)
\nonumber\\
=&\ - \frac{\kappa^2}{2}\int d^2T_t\,
\Big(\llangle\Psi_A\, (X_0\eta V_C)\, (b_0^- b_0^+) \Psi_B\, (\eta V_D)\rrangle_{W_t}
+\llangle\Psi_A\, (\eta V_C)\, (b_0^- b_0^+)\, \Psi_B\, (X_0\eta V_D)\rrangle_{W_t}\Big)
\nonumber\\
&\
-\frac{\kappa^2}{2}\oint d\theta_t\,
\Big(
\langle\Psi_A\, V_C (b_0^-\Xi)\Psi_B\,(\eta V_D)\rangle_{W_{t}}
+ \langle\Psi_A\,(\eta V_C) (b_0^-\Xi) \Psi_B\, V_D\rangle_{W_{t}}\Big)\Big|_{T_t=0}\,, 
\label{F2B2 t}
\end{align}
where we denoted $\oint d\theta\equiv\int^{2\pi}_0d\theta/2\pi$ for simplicity.
Here the first line of the final expression has the same form of
the external states as those in (\ref{F2B2 s}), but 
the extra contribution in the second line  
appears from the boundary $T_t=0$ by exchanging $Q$ and $b_0^+$ using
\begin{equation}
 \{Q, \int^\infty_0 dT\ b_0^+e^{-TL_0^+}\}\ =\
 \int^\infty_0 dT\ L_0^+e^{-TL_0^+}
\nonumber\\
=\ 1\,.
\end{equation}
The contribution from the $u$-channel has the same structure
and is calculated as
\begin{align}
 \mathcal{A}^{(AD|BC)}_{F^2B^2}\ =&\ -\kappa^2\int d^2T_u\,
\langle\Psi_A V_D (b_0^- b_0^+ X\eta) \Psi_B V_C\rangle_{W_u}
\nonumber\\
=&\ -\frac{\kappa^2}{2}\int d^2T_u\,
\left(\langle\Psi_A\, (QV_D) (b_0^- b_0^+ \Xi) \Psi_B\, (\eta V_C)\rangle_{W_u}
+\langle\Psi_A\, (\eta V_D) (b_0^- b_0^+ \Xi) \Psi_B\, (QV_C)\rangle_{W_u}\right)
\nonumber\\
&\
-\frac{\kappa^2}{2}\oint d\theta_u\,
\left(
\langle\Psi_A V_D (b_0^- \Xi)\Psi_B\,(\eta V_C)\rangle_{W_u}
+ \langle\Psi_A\,(\eta V_D) (b_0^- \Xi) \Psi_B V_C\rrangle_{W_u}\right)\Big|_{T_u=0}
\nonumber\\
=&\ - \frac{\kappa^2}{2}\int d^2T_u\,
\Big(\llangle\Psi_A\, (X_0\eta V_D) (b_0^- b_0^+) \Psi_B\, (\eta V_C)\rrangle_{W_u}
+\llangle\Psi_A\, (\eta V_D) (b_0^- b_0^+) \Psi_B\, (X_0\eta V_C)\rrangle_{W_u}\Big)
\nonumber\\
&\
-\frac{\kappa^2}{2}\oint d\theta_u\,
\left(
\langle\Psi_A V_D (b_0^- \Xi)\Psi_B\,(\eta V_C)\rangle_{W_u}
+ \langle\Psi_A\,(\eta V_D) (b_0^- \Xi) \Psi_B V_C\rangle_{W_u}\right)\Big|_{T_u=0}\,.
\end{align}

The contribution from the contact type diagram can be obtained
using the vertices (\ref{NS2R2}):
\begin{align}
 \mathcal{A}^{(ABCD)}_{F^2B^2}\ =&\ -\frac{\kappa^2}{2}\int d\theta_1 d\theta_2
\langle (\xi_0 b_{C_1} b_{C_2})
\Psi_A \Psi_B \big((QV_C)\, (\eta V_D) + (\eta V_C)\, (QV_D)\big)\rangle_{W_4}
\nonumber\\
&\
-\frac{\kappa^2}{2}\oint d\theta_t\,\Big(
\langle\Xi b_0^- \big(\Psi_A\, (\eta V_C)\big)\, \Psi_B\, V_D\rangle_{W_t}
\langle\Xi b_0^- \big(\Psi_B\, (\eta V_C)\big)\, \Psi_A\, V_D\rangle_{W_t}
\Big)\Big|_{T_t=0}
\nonumber\\
&\
-\frac{\kappa^2}{2}\oint d\theta_u\,\Big(
\langle\Xi b_0^- \big(\Psi_A (\eta V_D)\big)\, \Psi_B\, V_C\rangle_{W_u}
+\langle\Xi b_0^- \big(\Psi_B (\eta V_D)\big)\, \Psi_A\, V_C\rangle_{W_u}
\Big)\Big|_{T_u=0}
\nonumber\\
%
=&\ -\frac{\kappa^2}{2}\int d\theta_1 d\theta_2
\llangle (b_{C_1} b_{C_2}) 
 \Psi_A \Psi_B \big((X_0\eta V_C)\, (\eta V_D) 
+ (\eta V_C)\, (X_0\eta V_D)\big)\rrangle_{W_4}
\nonumber\\
&\
+\frac{\kappa^2}{2}\oint d\theta_t\,
\Big(
\langle\Psi_A\,(\eta V_C) (b_0^-\Xi) \Psi_B V_D\rangle_{W_t}
+\langle\Psi_A V_C (b_0^- \Xi) \Psi_B\, (\eta V_D)\rangle_{W_t}
\Big)\Big|_{T_t=0}
\nonumber\\
&\
+\frac{\kappa^2}{2}\oint d\theta_u\,\Big(
\langle\Psi_A\,(\eta V_D) (b_0^-\Xi) \Psi_B V_C\rangle_{W_u}
+\langle\Psi_A V_D (b_0^- \Xi) \Psi_B\, (\eta V_C)\rangle_{W_u}
\Big)\Big|_{T_u=0}\,.
\end{align}
The first line is the similar contribution to the one
in the four-fermion amplitude (\ref{four fermion contact}), coming from
the proper four-string interaction defined by 
the three-string product $[\,\cdot\,,\, \cdot\,,\, \cdot]$.
Extra contribution in the second line comes from the vertices represented by
the nested two string products $[\,\cdot\,,[\,\cdot\,,\,\cdot\,]]$,
represented by the diagram with the collapsed-propagator, the propagator with $T=0$, 
integrated only over its angle.
For the more details, see \cite{Kunitomo:2015hda}.

Summing up all these contributions, one can find that 
all the extra contributions integrated only one parameter are cancelled out.
The two-fermion-two-boson amplitude finally becomes
\begin{align}
 \mathcal{A}_{F^2B^2}\ =&\
\mathcal{A}_{F^2B^2}^{(AB|CD)}
+ \mathcal{A}_{F^2B^2}^{(AC|BD)}
+ \mathcal{A}_{F^2B^2}^{(AD|BC)}
+ \mathcal{A}_{F^2B^2}^{(ABCD)}
\nonumber\\
=&\
-\frac{\kappa^2}{2}\int d^2T_s\, \Big(
\llangle\Psi_A\Psi_B (b_0^- b_0^+) 
\big((X_0\eta V_C)\, (\eta V_D) + (\eta V_C)\, (X_0\eta V_D)\big)\rrangle_{W_s}
\Big)
\nonumber\\
&\ 
-\frac{\kappa^2}{2}\int d^2T_t\, \Big(
\llangle\Psi_A\, (X_0\eta V_C)\, (b_0^- b_0^+) \Psi_B\, (\eta V_D)\rrangle_{W_t}
+\llangle\Psi_A\, (\eta V_C)\, (b_0^- b_0^+)\, \Psi_B\, (X_0\eta V_D)\rrangle_{W_t}
\Big)
\nonumber\\
&\ 
-\frac{\kappa^2}{2}\int d^2T_u\, \Big(
\llangle\Psi_A\, (X_0\eta V_D) (b_0^- b_0^+) \Psi_B\, (\eta V_C)\rrangle_{W_u}
+\llangle\Psi_A\, (\eta V_D) (b_0^- b_0^+) \Psi_B\, (X_0\eta V_C)\rrangle_{W_u}
\Big)
\nonumber\\
&\
-\frac{\kappa^2}{2}\int d\theta_1 d\theta_2\
\llangle (b_{C_1} b_{C_2}) 
 \Psi_A \Psi_B 
\big((X_0\eta V_C)\, (\eta V_D) + (\eta V_C)\, (X_0\eta V_D)\big)\rrangle_{W_4}\,.
\end{align}
For the same reason as the four-fermion amplitude, this agrees with
the result in the first-quantized formulation. 
%

\end{document}